\documentclass[pdflatex,sn-nature]{sn-jnl}
\usepackage[utf8]{inputenc}
\usepackage[T1]{fontenc}
\usepackage{floatrow}
\usepackage{graphicx}%
\usepackage{multirow}%
\usepackage{amsmath,amssymb,amsfonts}%
\usepackage{amsthm}%
\usepackage{mathrsfs}%
\usepackage[title]{appendix}%
\usepackage{xcolor}%
\usepackage{textcomp}%
\usepackage{manyfoot}%
\usepackage{booktabs}%
\usepackage{algorithm}%
\usepackage{algorithmicx}%
\usepackage{algpseudocode}%
\usepackage{listings}%
\theoremstyle{thmstyleone}%
\usepackage{ragged2e}

\theoremstyle{thmstyletwo}%

\theoremstyle{thmstylethree}%

\usepackage{orcidlink}
\usepackage{physics}
\usepackage{bm}
\usepackage{mathtools}

\newcommand{\eqnref}[1]{Eq.~\eqref{#1}}

\newcommand{\figref}[1]{Fig.~\ref{#1}}
\newcommand{\subfigref}[2]{Fig.~\hyperref[#1]{\ref*{#1}(#2)}}
\newcommand{\subfigsref}[3]{Figs.~\hyperref[#1]{\ref*{#1}(#2)}-\hyperref[#1]{\ref*{fig:#1}(#3)}}

\newcommand{\mr}[1]{\mathrm{#1}}
\newcommand{\hc}{\mr{H.c.}}

\usepackage[low-sup]{subdepth}

\usepackage[]{newtxtext} 
\usepackage[subscriptcorrection,nosymbolsc,smallerops,bigdelims]{newtxmath} 
\DeclareMathAlphabet{\mathcal}{OMS}{cmsy}{m}{n} 
\DeclareMathAlphabet{\mathbcal}{OMS}{cmsy}{b}{n} 

\usepackage[activate={true,nocompatibility},final,tracking=alltext,kerning=true,spacing=true,protrusion=true,factor=1050,stretch=5,shrink=7,selected=true ,letterspace=-0]{microtype}

\begin{document}

\title{Hybrid acousto-optical swing-up state control in a quantum dot}

\author[]{\fnm{Mateusz} \sur{Kuniej}\,\orcidlink{0000-0001-5476-4856}}\email{mateusz.kuniej@pwr.edu.pl}

\author[]{\fnm{Pawe{\l}} \sur{Machnikowski}\,\orcidlink{0000-0003-0349-1725}}\email{pawel.machnikowski@pwr.edu.pl}

\author*[]{\fnm{Micha{\l}} \sur{Gawe{\l}czyk}\,\orcidlink{0000-0003-2299-140X}}\email{michal.gawelczyk@pwr.edu.pl}

\affil[]{\orgdiv{Institute of Theoretical Physics}, \orgname{Wroc\l{}aw University of Science and Technology}, \orgaddress{\street{Wybrzeże Stanisława Wyspiańskiego 27}, \city{Wroc\l{}aw}, \postcode{50-370}, \country{Poland}}}

\abstract{State transfer between different quantum systems is key for successful quantum technologies. Over long distances, photons are irreplaceable, but on short ranges in miniaturized complex devices or hybrid systems, coupling via orders of magnitude shorter-wavelength acoustic waves has great potential.
With interfaces to light, acoustic waves, and more, optically active quantum dots (QDs) are essential for multi-component systems. Here, we propose a hybrid acousto-optical method for non-resonant QD charge state control, extending the recent all-optical swing-up state preparation. We show that exciton and biexciton states, or other superpositions of charge states, can be prepared. Each field can act as a trigger, allowing for the implementation of either an optically gated acoustic control or the opposite scheme, where an optical pulse controls the transition during acoustic modulation. Thus, we introduce acoustic state control into a system that lacks direct acoustic coupling between the states. The method does not rely on pulse shaping and is expected to work with arbitrary pulse shapes as long as the optical dressing is performed quasi-adiabatically. Evaluating the phonon impact, we find an almost decoherence-free exciton preparation even at elevated temperatures with current QD and acoustic technology. This approach may also pave the way for optically controlled entanglement between emitters and acoustic modes, and further on-chip state transfer via quantum acoustic buses.}


\maketitle

\section*{Introduction}

Quantum information science and engineering have at their disposal many specialized nanoscopic systems \cite{Alfieri2022}, and the success of the second quantum revolution will depend on the effective use of their combined and structured layouts.
Thus, in addition to developing individual solid-state quantum systems, it becomes important to explore short- and long-distance coupling and state transmission between different types of them.
For long-haul quantum transmission, flying qubits, photons, and their entangled pairs are a natural choice. On the other hand, for in-place quantum computation and data processing, quantum-enhanced sensing, or quantum memories and repeaters, a common coupling connecting all the involved systems is essential and can be provided by augmenting optical protocols with either acoustic waves or purely acoustically.

Recent years have brought phononic structures \cite{Roth2023, Preuss2022} and coherent acoustic waves for quantum technologies. Surface acoustic waves (SAWs) with microwave frequencies are well established \cite{Choquer2022, Wang2018} and methods for generating THz coherent acoustic waves were also developed \cite{Maznev2012, Barajas-Aguilar2024, Akimov2015, Ruello2015}.
Acoustic waves have already found applications in radio signal processing, acousto-fluidics, life sciences, and biological and chemical sensors \cite{Delsing2019}. In quantum technologies, they have been used for spin-qubit control in color centers \cite{Clark2024, Dietz2023}, shuttling electrons between gate-defined QDs \cite{Jadot2021}, superconducting qubit-cavity coupling \cite{Moores2018, Sletten2019}, and coherent control in van der Waals materials \cite{Lazić2019, Preuss2022}. The naturally short acoustic wavelengths and universal mechanical coupling to all solid-state systems open the way to the miniaturization of future quantum devices \cite{Delsing2019}, especially multi-component quantum hybrid systems coupled via a universal acoustic quantum bus, and combining different degrees of freedom to realize quantum information transfer, processing, storage, communication, or quantum-enhanced detection \cite{Pezze2021, Gu2023, Thomas2024}.

Optically active semiconductor QDs are still at the forefront of platforms for quantum technologies. They offer extensive tunability of optical properties and cover a wide range of energy scales related to QD charge and spin qubits, with applications in quantum communication, metrology, and information processing \cite{Vajner2022, Michler2017, Benyoucef2023}. Above all, QDs can operate as very efficient and high-quality quantum emitters \cite{Kim2020, Meng2022}, creating polarization-entangled photon pairs \cite{Vajner2022, Hafenbrak2007}. Consequently, due to their compatibility with light, microwaves, and mechanical waves, QDs represent an excellent component of hybrid systems \cite{Clerk2020, Gu2023, Kim2022}. Although their coupling with deformations is limited to shifting state energies, i.e., phonons cannot create electron-hole pair excitations, such phenomena as optomechanical wave mixing or modulated resonance fluorescence have been studied \cite{Weiss2021, Wigger2021}.

Various methods are used to prepare charge states in QDs: resonant excitation \cite{Stievater2001, Kamada2001}, phonon-assisted schemes \cite{Thomas2021, Vyvlecka2023}, adiabatic rapid passage \cite{Simon2011, Kaldeway2017}, two-photon excitation \cite{Stufler2006, Machnikowski2008}, or schemes using resonant component frequencies \cite{He2019, Koong2021}. However, each method has specific limitations, either needing spectral \cite{He2019, Koong2021} or polarization filtering \cite{Matthiesen2012} or being incoherent \cite{Ardelt2014, Barth2016}. Therefore, there is a long-standing demand for coherent non-resonant state preparation. Recently, a new approach has been proposed \cite{Bracht2021, Bracht2022, Bracht2023} and realized \cite{Karli2022, Boos2024}. It involves two off-resonant optical pulses, which, due to their beating behavior, effectively modulate the system's eigenenergies, in turn leading to an excitation with a swing-up behavior. The method is innovative and promising, as it is non-resonant and provides unmatched quality of the state preparation.

Here, we extend the swing-up method to a hybrid acousto-optical state-control scheme, aiming to meet the demand for new off-resonant excitation schemes and overcome the natural limitation of the diagonal coupling of QDs to material deformations. We propose to leverage recent progress in acoustic control of quantum states and optical processes and directly modulate the detuning by exploiting the deformation-potential coupling of the QD-confined carriers with an acoustic wave. Our method engages phonons in state control, although they could not induce it independently. In addition to enabling optically gated acoustic control of a QD-defined qubit that normally lacks such transitions, this approach may pave the way for optically controlled entanglement between emitters and acoustic modes and for further state transfer, leading to acoustically connected multi-component devices. We describe this scheme theoretically and show that either optical or acoustic fields can be treated as the ones controlling the state evolution. By directly calculating the system evolution, we show how both the exciton and biexciton states can be deterministically prepared. The method does not rely on pulse shaping, and we expect it to work for any pulse shapes as long as optical dressing is quasi-adiabatic. In our considerations, we use basic shapes for the sake of simplicity and to obtain analytical results where possible. Starting with flat-top pulses, we present analytical expressions for the required acoustic field parameters. For more realistic Gaussian pulses, we show how to optimize the protocol around those estimated parameters to achieve full occupation of the desired state. Finally, we estimate the phonon-induced decoherence for two relevant types of epitaxial QDs: standard self-assembled (Stranski-Krastanov) InAs/GaAs dots and GaAs/AlGaAs QDs grown by droplet etching epitaxy \cite{daSilva2021}. We find that our scheme is almost decoherence-free for GaAs/AlGaAs QDs even at elevated temperatures, even for the sub-THz acoustic regime. Further increasing the detuning (and consequently, the required acoustic field frequency) to the THz range can further improve the fidelity, which indicates the importance of developing THz acoustic technology.

In the following, we define the system and formulate its theoretical description, and then in the first part of the analysis, we work in a simplified model to gain some understanding and find approximate conditions for the acoustic field needed to prepare the exciton. Then, we deal with biexciton preparation in the full model. Finally, we evaluate the fidelity of all prepared states.

\section*{Results}\label{sec:results}

\subsection*{System and its model}\label{sec:system}

\begin{figure}[tbp!]
    \centering
    \includegraphics[width=0.5\columnwidth]{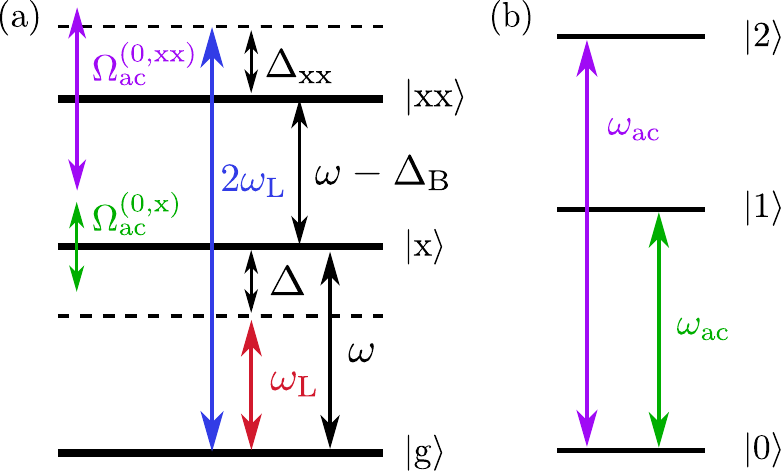}
    \caption{\textbf{Schematic energy structure of the studied three-level system with external fields.} (a) Ground $\ket{\mathrm{g}}$ and two excited levels, $\ket{\mathrm{x}}$ and $\ket{\mathrm{xx}}$, corresponding to the exciton and biexciton in a QD. Acoustic modulation of the excited states is marked with green and violet arrows. Two optical driving cases are marked: $\ket{\mathrm{x}}$ preparation (red arrow) and $\ket{\mathrm{xx}}$ preparation (blue arrow); (b) Dressed-states picture with acoustic field frequencies tuned for $\ket{\mathrm{g}}$--$\ket{\mathrm{x}}$ (green arrow) and $\ket{\mathrm{g}}$--$\ket{\mathrm{xx}}$ (violet arrow) transitions.}
    \label{fig:system}
\end{figure}

We consider a QD coupled to a linearly polarized laser detuned from the optical transitions of interest. According to the selection rules, an appropriate linear polarization allows for single-photon transitions from the empty dot ($\ket{\mathrm{g}}$) to the selected exciton ($\ket{\mathrm{x}}$) and from exciton to biexciton ($\ket{\mathrm{xx}}$) states. The system thus has an effective three-level structure, as shown in \subfigref{fig:system}{a}. The exciton energy is $\hbar\omega$. The biexciton state can be approximately described as a pair of excitons. Due to interactions, its energy is lower than $2\hbar\omega$ by the binding energy $\Delta_{\mathrm{B}}$, for which we assume a typical value of 4~meV. The red and blue arrows mark the two detuned optical couplings: $\ket{\mathrm{g}}$-$\ket{\mathrm{x}}$ and two-photon $\ket{\mathrm{g}}$-$\ket{\mathrm{xx}}$. Additionally, the QD is modulated by an external acoustic field (green and violet arrows marking the modulation of the levels).

We will use two- and three-level system models, where the simpler one will provide us with some intuitive insight, while the final quantitative results will come from the three-level model. Let us start with two-level systems, where the acoustic field couples only to the exciton state. The Hamiltonian can be written as
\begin{equation}
    H(t) = \hbar\omega\ketbra{\mathrm{x}}{\mathrm{x}} + \hbar\Omega_{\mathrm{ac}}(t)\ketbra{\mathrm{x}}{\mathrm{x}} - \bm{d}\cdot\bm{E}(t),
\end{equation}
where $\textit{\textbf{E}}(t) = \textit{\textbf{E}}_0(t)\cos(\omega_{\mathrm{L}}t)$ is the classical optical field, $\bm{d}$ is the dipole moment operator describing the coupling to light. We assume $\bra{\mathrm{g}}\bm{d}\ket{\mathrm{g}} = \bra{\mathrm{x}}\bm{d}\ket{\mathrm{x}} = 0$. Further, $\Omega_{\mathrm{ac}}(t) = \Omega_{\mathrm{ac}}^{(0)}(t)\cos(\omega_{\mathrm{ac}}t)$ is the acoustic field, with a slowly varying envelope $\Omega_{\mathrm{ac}}^{(0)}(t)$. This Hamiltonian in the rotating wave approximation (RWA) and rotating frame takes the form
\begin{equation}
    \begin{split}
        H(t) =& -\hbar\Delta\ketbra{\mathrm{x}}{\mathrm{x}} + \hbar\Omega_{\mathrm{ac}}(t)\ketbra{\mathrm{x}}{\mathrm{x}} \\ &+ \frac{1}{2}\hbar\Omega_{\mathrm{L}}^{(0)}(t)\left(\ketbra{\mathrm{g}}{\mathrm{x}} + \hc\right) \\ =& H_0 + H_{\mathrm{ac}}(t) + H_{\mathrm{L}}(t),
    \end{split}
    \label{eq:twoLevelSystemHamiltonian}
\end{equation}
where $\Delta = \omega_{\mathrm{L}} - \omega$ is the laser detuning.

In a three-level model including the biexciton state, $\ket{\mathrm{xx}}$, the acoustic field modulates both excited states with different amplitudes due to different coupling strengths. There is no first-order (one-photon) optical coupling between the $\ket{\mathrm{g}}$ and $\ket{\mathrm{xx}}$ states, but they can be coupled by two-photon transitions. The three-level system Hamiltonian in the RWA (Supplementary Note 1\cite{supplement} for derivation) can be written in the form
\begin{equation}
    \begin{split}
        \!\!\!H(t) = &-\hbar\Delta\ketbra{\mathrm{x}}{\mathrm{x}} - \hbar\Delta_{\mathrm{xx}}\ketbra{\mathrm{xx}}{\mathrm{xx}} \\ & + \frac{1}{2}\hbar\Omega_{\mathrm{L}}^{(0)}(t)(\ketbra{\mathrm{x}}{\mathrm{g}} + \ketbra{\mathrm{xx}}{\mathrm{x}} + \hc) \\ & + \hbar\cos(\omega_{\mathrm{ac}}t)\left(\Omega_{\mathrm{ac}}^{(0,\mathrm{x})}(t)\ketbra{\mathrm{x}}{\mathrm{x}} + \Omega_{\mathrm{ac}}^{(0,\mathrm{xx})}(t)\ketbra{\mathrm{xx}}{\mathrm{xx}}\right),
    \end{split}
    \label{eq:threeLevelSystemHamiltonian}
\end{equation}
where $\Omega_{\mathrm{ac}}^{(0,\mathrm{x})}(t)$ and $\Omega_{\mathrm{ac}}^{(0,\mathrm{xx})}(t)$ are the acoustic coupling amplitudes for the two excited states, and $\Delta_{\mathrm{xx}} = 2\Delta + \Delta_{\mathrm{B}}$ is the detuning from the two-photon transition to the $\ket{\mathrm{xx}}$ state. With such definitions, negative $\Delta$ or $\Delta_{\mathrm{xx}}$ means blue detuning, and the opposite sign means red detuning with respect to either one- or two-photon transition. To estimate the strength of the phonon-biexciton coupling, we assume the biexciton to be composed of two uncorrelated excitons.
Then, we have
\begin{equation}
    \bra{\mathrm{xx}}H_{\mathrm{ac}}\ket{\mathrm{xx}} \approx 2\bra{\mathrm{x}}H_{\mathrm{ac}}\ket{\mathrm{x}},
\end{equation}
where $H_{\mathrm{ac}}$ is the carrier-deformation coupling Hamiltonian, i.e., the biexciton state is modulated approximately twice as strongly as the exciton.

We aim at obtaining the evolution leading to the full occupation of the $\ket{\mathrm{x}}$ [laser marked with the red arrow in \subfigref{fig:system}{a}] or $\ket{\mathrm{xx}}$ (blue arrow) state under off-resonant laser driving and with appropriately tuned acoustic modulation.
Thanks to sinusoidally varying detunings resulting from the acoustic modulation of both excited states' energies, we should obtain the swing-up behavior when the acoustic frequency matches the respective Rabi frequencies for the system driven by the laser alone, in analogy to the original swing-up concept \cite{Bracht2021}. Those Rabi frequencies correspond to the splittings between the new eigenstates (dressed states) that one can use to describe the laser-driven system, schematically shown in \subfigref{fig:system}{b} and discussed later. The two arrows show the acoustic frequencies that will lead to the occupation of the $\ket{\mathrm{x}}$ (green arrow) and $\ket{\mathrm{xx}}$ (violet) states.

In this and most of the next subsections, we will study the evolution of the closed system by directly solving the Liouville-von~Neumann equation
\begin{equation}\label{eq:LvN}
    \Dot{\rho}(t) = -\frac{i}{\hbar}\left[H(t), \rho(t)\right],
\end{equation}
where $\rho(t)$ is the density matrix, with the initial condition given by $\rho(t_0) = \ketbra{\mathrm{g}}{\mathrm{g}}$, i.e., the system is initially prepared in the ground state. The numerical solution of \eqnref{eq:LvN} gives us the exact evolution of a closed system.
However, the prepared states undergo recombination, which may diminish their obtained occupation, and even in high-quality QD samples at low temperatures, carriers trapped in QDs couple to phonons, which can cause decoherence. We address these issues separately in the last subsection, where we account for recombination by solving the Lindblad equation
\begin{equation}\label{eq:Lindblad}
    \Dot{\rho}(t) = -\frac{i}{\hbar}\left[H(t), \rho(t)\right] + \sum_{\mu = \mathrm{x, xx}} \gamma_{\mu}\left(L_{\mu}\rho(t)L^{\dagger}_{\mu} - \frac{1}{2}\left\{L^{\dagger}_{\mu}L_{\mu}, \rho(t)\right\} \right),
\end{equation}
where $L_{\mathrm{x}} = \ketbra{\mathrm{g}}{\mathrm{x}}$, $L_{\mathrm{xx}} = \ketbra{\mathrm{x}}{\mathrm{xx}}$, $\gamma_{\mathrm{x}}$ and $\gamma_{\mathrm{xx}}$ are the exciton and biexciton recombination rates, and separately, we include the deformation-potential and piezoelectric charge-phonon couplings to study the decoherence due to the dynamical phonon reaction to the evolution of the system in a non-Markovian approach. In this way, we evaluate the achievable fidelity values for states prepared with the acousto-optical scheme.

\subsection*{Two-level system: exciton preparation}\label{sec:two-level}
While we will finally evaluate the state preparation fidelity in a three-level system for both $\ket{\mathrm{x}}$ and $\ket{\mathrm{xx}}$ states to keep it realistic, in this Section, we begin the analysis with a two-level system to provide some intuition and analytical estimates for the needed acoustic field characteristics.

\subsubsection*{Resonance condition for acousto-optical control}
In the case of a two-level system described by the Hamiltonian~\eqref{eq:twoLevelSystemHamiltonian}, we can determine the conditions leading to the desired evolution analytically. To find the required acoustic field parameters, we first diagonalize the Hamiltonian $H_0 + H_{\mathrm{L}}$, i.e., of the system coupled to the laser only. For slowly varying laser pulse envelopes, the transformation is performed at each point in time and yields
\begin{equation}
    H_0 + H_{\mathrm{L}}(t) = E_+(t)\ketbra{+}{+} + E_-(t)\ketbra{-}{-},
\end{equation}
where the time-dependent energy difference 
\begin{equation}
     E_+(t) - E_-(t) = \hbar\sqrt{\Delta^2 + \left[\Omega_{\mathrm{L}}^{(0)}(t)\right]^2} \equiv\hbar\Omega_{\mathrm{R}}(t)
     \label{eq:RabiFrequency}
\end{equation}
is now the Rabi frequency of the system described in the instantaneous basis of states dressed by the light,
\begin{equation}
    \begin{split}
        \ket{+(t)} & = \sin\frac{\theta(t)}{2}\ket{\mathrm{g}} + \cos\frac{\theta(t)}{2}\ket{\mathrm{x}},\\
        \ket{-(t)} & = \cos\frac{\theta(t)}{2}\ket{\mathrm{g}} - \sin\frac{\theta(t)}{2}\ket{\mathrm{x}},
    \end{split}
\end{equation}
where the time-dependent mixing angle $\theta(t)$ is given by
\begin{equation}
    \theta(t) = \tan^{-1}\left(\frac{\Omega_{\mathrm{L}}^{(0)}(t)}{\Delta}\right).
\end{equation}
The acoustic field Hamiltonian now takes the form
\begin{equation}
    \begin{split}
        H_{\mathrm{ac}}(t) = &\ \hbar\Omega_{\mathrm{ac}}(t)\left(\sin^2\frac{\theta(t)}{2}\ketbra{+}{+} + \cos^2\frac{\theta(t)}{2}\ketbra{-}{-}\right) \\ & + \frac{1}{2}\hbar\Omega_{\mathrm{ac}}(t)\sin\theta(t)\left(\ketbra{+}{-} + \ketbra{-}{+}\right).
    \end{split}
    \label{eq:acousticHamiltonianTwoLevel}
\end{equation}
Having both diagonal and off-diagonal elements, $H_{\mathrm{ac}}$ leads to energy shifts
\begin{subequations}
\begin{align}
 E_+(t) \rightarrow E_{+,\mathrm{ac}}(t) &= E_+(t) + \hbar\Omega_{\mathrm{ac}}(t)\sin^2\frac{\theta(t)}{2} \\
 E_-(t) \rightarrow E_{-,\mathrm{ac}}(t) &= E_-(t) + \hbar\Omega_{\mathrm{ac}}(t)\cos^2\frac{\theta(t)}{2}
\end{align}
\end{subequations}
and, more importantly, transitions between the dressed states. The evolution is thus driven by the off-diagonal terms $\propto \Omega_{\mathrm{ac}}(t) \propto \exp(\pm i\omega_{\mathrm{ac}}t)$. To obtain the resonance condition, the frequency $\omega_{\mathrm{ac}}$ has to correspond to the energy difference between the dressed states. However, this difference depends on time and $\omega_{\mathrm{ac}}$ itself, leading to a generally nontrivial criterion. To find a practically feasible condition, we neglect fast oscillations of the dressed states' energies and take their mean energy difference as in the original, all-optical swing-up protocol \cite{Bracht2023}. The resonance condition is then given by
\begin{equation}\label{eq:omega-ac}
    \omega_{\mathrm{ac}}(t) = \Omega_{\mathrm{R}}(t). 
\end{equation}
This condition will hold exactly for flat-top pulses when $\Omega_{\mathrm{R}}$ is constant during the pulse plateau, while some additional optimization around this point will be needed for realistic, e.g., Gaussian pulses.

\subsubsection*{Acoustic control during optical driving}
We can continue the fully analytical discussion for the case of flat-top pulses with the acoustic pulse shorter than the optical one. This situation also corresponds to quasi-continuous optical driving with an acoustic pulse that triggers evolution. We model flat-top pulse envelopes with
\begin{equation}\label{eq:flat-top}
    \Omega_{\eta}^{(0)}(t) = \frac{1}{2}A_{\eta}\left[1-\erf\left(\frac{t-\sigma_{\eta}}{\kappa_{\eta}}\right)\erf\left(\frac{t+\sigma_{\eta}}{\kappa_{\eta}}\right)\right],
\end{equation}
where $A_{\eta}$ is the amplitude, for both optical ($\eta=\mathrm{L}$) and acoustic ($\eta=\mathrm{ac}$) fields. Such a pulse has a plateau of duration $\sim\sigma_{\eta}$, and the switching rate is controlled by the parameter $\kappa$.

\begin{figure}[tbp!]
    \centering
    \includegraphics[width=0.5\columnwidth]{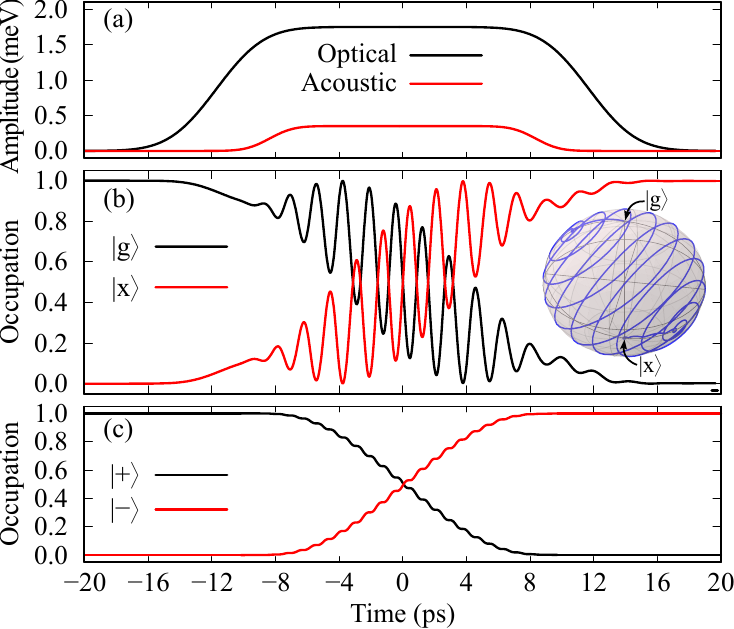}
    \caption{\textbf{Optically gated acoustic control.} Exciton preparation with quasi-continuous optical coupling and acoustic $\pi$-rotation control pulse: (a) envelopes of acoustic and optical pulses, (b) bare states occupations; inset shows state evolution on the Bloch sphere, (c) occupations of the dressed states. Parameters used: $\hbar\Delta = 1.75$~meV, $\hbar A_{\mathrm{L}} = 1.75$~meV, $\sigma_{\mathrm{L}} = 11.7$~ps, $\hbar A_{\mathrm{ac}} = 0.35$~meV, $\sigma_{\mathrm{ac}} = 8.36$~ps, $\hbar\omega_{\mathrm{ac}} = 2.474$~meV, $\kappa_{\mathrm{L}} = 3.34$~ps, and $\kappa_{\mathrm{ac}} = 1.67$~ps.}
    \label{fig:OccupationConstField}
\end{figure}

The parameter space to explore is large. We choose to fix the laser amplitude $A_{\mathrm{L}}$ equal to the detuning $\Delta$ so that we do not enter a regime where the Rabi frequency is dominated by one of these parameters. Doing so, we avoid negligible detunings and situations corresponding to the trivial addition of laser and acoustic frequencies. Let $\hbar\Delta = 1.75$~meV, a typical value providing enough spectral separation between the laser and the transition in a QD. For such a case, the needed acoustic field frequency $\hbar\omega_{\mathrm{ac}}\sim 2.474$~meV is in the range that is currently experimentally available \cite{Akimov2015, Ruello2015}. In this scenario, we want the optical pulse to be long enough for the acoustic one to operate during the optical plateau. We then have a fixed criterion for the acoustic frequency, $\omega_{\mathrm{ac}}=\Omega_{\mathrm{R}} = \sqrt{\cramped{\Delta^2}+\cramped{A_{\mathrm{L}}^2}}$. It is reasonable to assume that for the protocol to work properly, the acoustic pulse should be long enough for the energy of the oscillating states to average out, i.e., at least a few periods ($\tau_{\mathrm{ac}} = 2\pi/\Omega_{\mathrm{ac}}$) of the acoustic oscillation (for most of the paper, we keep it $\gtrsim 5\tau_{\mathrm{ac}}$). We will, however, show that the method works even beyond that limit. Lastly, the optical dressing and undressing of states must be performed quasi-adiabatically. We choose $\sigma_{\mathrm{ac}} = 5 \tau_{\mathrm{ac}}$, $\kappa_{\mathrm{L}} = 2\tau_{\mathrm{ac}}$, $\kappa_{\mathrm{ac}} = \tau_{\mathrm{ac}}$ and $\sigma_{\mathrm{L}} = 7\tau_{\mathrm{ac}}$. The corresponding pulse envelopes are shown in \subfigref{fig:OccupationConstField}{a}.

Now, we can fix the amplitude of the acoustic pulse. We want it to cause a $\pi$ rotation in the dressed-state basis, from the $\ket{+}$ to the $\ket{-}$ state. If the entire acoustic pulse occurs during the plateau of the optical field envelope, we can analytically determine the required acoustic pulse amplitude. In such a situation, the mixing angle $\theta$ [\eqnref{eq:acousticHamiltonianTwoLevel}] is time-independent, and the effective driving amplitude is $\hbar A_{\mathrm{ac}}\sin\theta/2$ [cf.~\eqnref{eq:flat-top}]. Thus, for a $\pi$ rotation we need
\begin{equation}
    A_{\mathrm{ac}} = \frac{\pi}{\sigma_{\mathrm{ac}}\sin\theta} =  \frac{\pi}{\sigma_{\mathrm{ac}}}\sqrt{1+\frac{\Delta^2}{A_{\mathrm{L}}^2}},
    \label{eq:acousticAmplitudeCondition}
\end{equation}
which for our choice of $A_{\mathrm{L}}=\Delta$ gives $A_{\mathrm{ac}} = \sqrt{2}\pi/\sigma_{\mathrm{ac}}$. 

We show the result of a numerical simulation for such a pulse in \subfigref{fig:OccupationConstField}{b} where the system initially prepared in the $\ket{\mathrm{g}}$ state is completely transferred into the state $\ket{\mathrm{x}}$. The inset shows the swing-up state evolution on the Bloch sphere. As mentioned, the acoustic field envelope satisfies the condition given in \eqnref{eq:acousticAmplitudeCondition} and is switched on only during the optical pulse plateau, as shown in \subfigref{fig:OccupationConstField}{a}. Thus, it acts only on the states that are already dressed by light, modulating the rotation axis of the qubit periodically. \subfigref{fig:OccupationConstField}{c} additionally presents the occupation of the dressed states during the evolution. As predicted, the transition is triggered only by the acoustic pulse, which completely transfers the occupation. During the switch-off process of the optical field, the state $\ket{-}$ quasi-adiabatically undresses to the excited state.
Note that for the chosen parameters, we have the ratio $A_{\mathrm{ac}}/\omega_{\mathrm{ac}}\sim0.1$, which corresponds to weak modulation under which no spectral shift of the QD transition is expected and less than 1\% of emission will redistribute to the sidebands \cite{Wigger2021}. We maintain this regime for most of the discussion in the paper.

\subsubsection*{Optically driven transition during acoustic modulation}
For a more realistic simulation and to explore the case with the reversed role of the fields (optical pulse as a trigger), we now choose a continuous-wave acoustic field and a Gaussian optical pulse,
\begin{equation}
    \Omega_{\mathrm{ac}}^{(0)} = A_{\mathrm{ac}},\quad \Omega_{\mathrm{L}}^{(0)}(t) = A_{\mathrm{L}}\exp\left(-\frac{t^2}{2\sigma_{\mathrm{L}}^2}\right).
\label{eq:envelopes}
\end{equation}
One should remember that the dressing and undressing procedure has to be adiabatic to fully invert the system occupation. Thus, the laser pulse cannot be too short. We maintain the parameters as in the previous subsection. This case has a far-reaching application potential, which may allow entangling QD states with an acoustic mode and further the transfer of a quantum state via an acoustic bus.

\begin{figure}[tbp!]
    \centering
    \includegraphics[width=0.5\columnwidth]{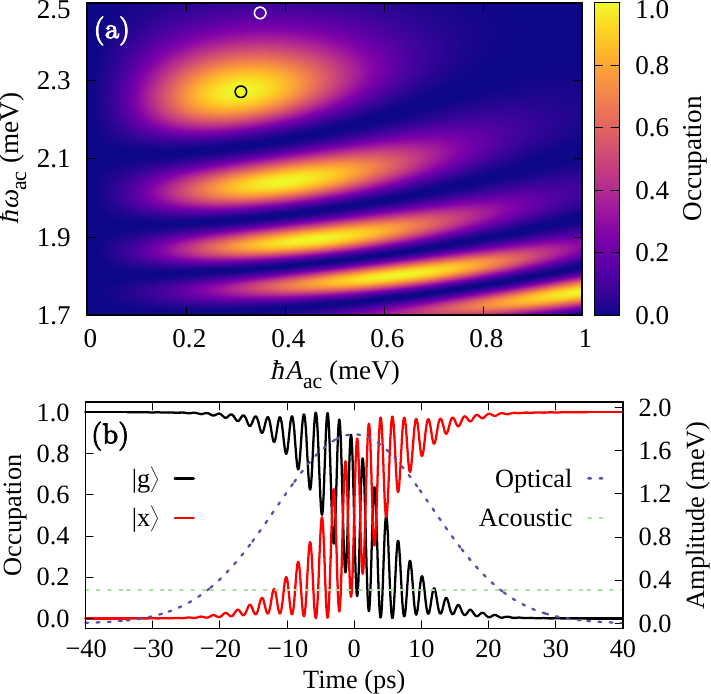}
    \caption{\textbf{Acoustically gated optical control: exciton preparation}. \textbf{(a)} Optimal acoustic-field parameters for exciton preparation. Occupation of the exciton state $\ket{\mathrm{x}}$ as a function of acoustic field parameters for a Gaussian optical pulse ($\hbar\Delta = 1.75$~meV, $\hbar A_{\mathrm{L}} = 1.75$~meV, $\sigma_{\mathrm{L}} = 11.7$~ps) and constant envelope of acoustic driving. We mark a point corresponding to the maximum occupation (black circle) and a point given by analytical prediction [Eq.~\eqref{eq:omega-ac} and Eq.~\eqref{eq:acousticAmplitudeCondition}] for flat-top pulses (white circle).
    \textbf{(b)} Evolution of bare states occupation in the two-level system (left axis) and external field amplitudes (right axis) for acoustic field parameters corresponding to the black circle in panel (a).}
    \label{fig:excitonMap}
\end{figure}

The evolution in this scenario is more complex, and some corrections to our analytical predictions from \eqnref{eq:omega-ac} and \eqnref{eq:acousticAmplitudeCondition} are needed. Thus, we resort to numerical simulations in which we vary $\omega_{\mathrm{ac}}$ and $A_{\mathrm{ac}}$. We again choose $\sigma_{\mathrm{L}} = 11.7$~ps to obtain approximately the same protocol duration.
Fig.~\ref{fig:excitonMap}(a) shows the map of the resultant occupation of the state $\ket{\mathrm{x}}$. We achieve a full $\pi$-rotation for $\hbar\omega_{\mathrm{ac}}\approx 2.272$~meV and $\hbar A_{\mathrm{ac}} \approx 0.311$~meV (black circle), which are relatively close in terms of the phonon energy and amplitude to those from \eqnref{eq:omega-ac} and \eqnref{eq:acousticAmplitudeCondition} for flat-top pulses (white circle), and allow reasonable estimation of the needed parameters in the experiment. The small differences are mainly due to the overlap of the acoustic field with the periods of dressing and undressing of the states with the optical field, since the Gaussian optical pulse does not produce a constant energy splitting between the dressed states.

The other visible maxima in \figref{fig:excitonMap}(a) correspond to $3\pi$, $5\pi$, etc. rotations. The shift in the energy is once again caused by the impact of optical pulse tails overlapping with the acoustic field, which is more pronounced for stronger acoustic fields.

In Fig.~\ref{fig:excitonMap}(b), we present the evolution of bare states occupation (solid lines) for the optimal $\omega_{\mathrm{ac}}$ and $A_{\mathrm{ac}}$ from Fig.~\ref{fig:excitonMap}(a), plotted together with the envelopes of the continuous-wave acoustic and Gaussian-pulsed optical fields (dashed lines). One may notice the typical swing-up behavior ending with full occupation inversion.

\subsection*{Three-level system and biexciton preparation}\label{sec:three-level}
We now switch to a more realistic three-level model and focus on the biexciton preparation. We start our calculations with Eq.~\eqref{eq:threeLevelSystemHamiltonian}. Although there is no one-photon optical transition between the ground and biexciton states, we can construct an acousto-optical protocol that allows biexciton preparation, again in an analogy to the all-optical swing-up scheme \cite{Bracht2023}. As previously, we diagonalize the Hamiltonian without the acoustic field, which gives
\begin{equation}
    H(t) = E_0(t)\ketbra{0}{0} + E_1(t)\ketbra{1}{1} + E_2(t)\ketbra{2}{2}.
\end{equation}
written in the time-dependent dressed states
\begin{equation}
    \ket{n(t)} = \!\!\!\sum_{k\in\{\mathrm{g}, \mathrm{x}, \mathrm{xx}\}}\!\!\! a_{n,k}(t)\ket{k},
\end{equation}
with time-dependent coefficients $a_{n,k}(t)$ dependent on the optical field amplitude and the detuning.

Following a similar procedure as previously, we add the acoustic field Hamiltonian $H_{\mathrm{ac}}$, which also modulates the biexciton state. Writing $H_{\mathrm{ac}}$ in the basis of the dressed states $\ket{n(t)}$ results again in energy shifts and driving due to the off-diagonal elements $\propto\ketbra{n(t)}{m(t)}$. The resonance condition leading to oscillations between a pair of dressed states is given by
\begin{equation}
    \hbar\omega_{\mathrm{ac}}^{(m\leftrightarrow n)}(t) = E_n(t) - E_m(t).
\end{equation}
The new eigenstates of the three-level system are not easily analytically handled. Thus, we find the acoustic field parameters $\omega_{\mathrm{ac}}$ and $A_{\mathrm{ac}}$, needed for the $\pi$-rotation, via numerical optimization.

\begin{figure}[tbp!]
    \centering
    \includegraphics[width=0.5\columnwidth]{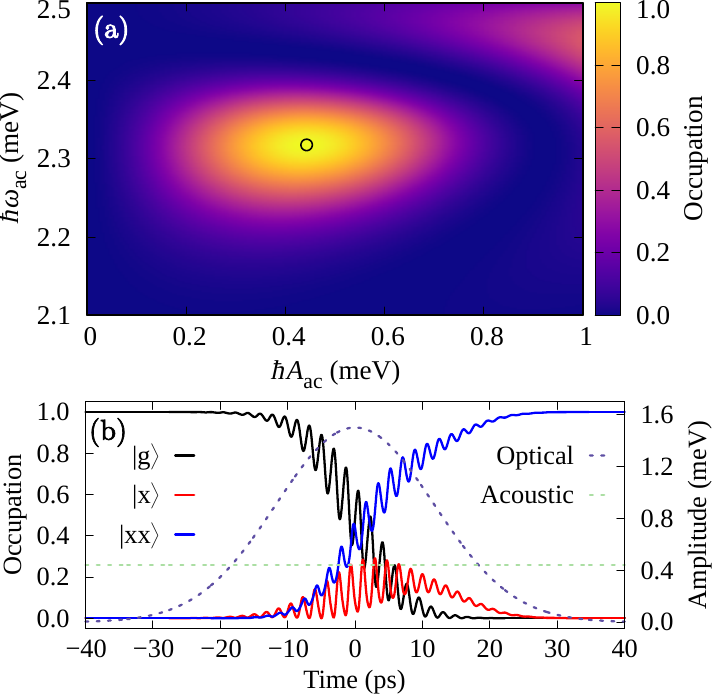}
    \caption{\textbf{Biexciton preparation.} \textbf{(a)} Occupation of the biexciton as a function of acoustic field parameters for a Gaussian optical pulse with ($\hbar\Delta_{\mathrm{xx}} = -2.5$~meV, $\hbar A_{\mathrm{L}} = 1.5$~meV, $\sigma_{\mathrm{L}} = 11.7$~ps) and constant envelope of acoustic driving. The black circle marks the optimal parameters.
    \textbf{(b)} Evolution of state occupations in the three-level system (left axis) and external field amplitudes (right) for acoustic field parameters corresponding to the black circle in panel (a).}
    \label{fig:biexcitonMap}
\end{figure}

We again consider a non-resonant Gaussian optical pulse during a constant acoustic driving [\eqnref{eq:envelopes}]. In Fig.~\ref{fig:biexcitonMap}(a), we scan the acoustic field frequency and amplitude to find the optimal parameters. The higher, $(2n+1)\pi$ rotations this time occur for significantly larger acoustic field amplitudes than in the exciton preparation protocol, and we do not show them here. The optimal parameters marked with the black circle are $\hbar\omega_{\mathrm{ac}} = 2.318$~meV, and $\hbar A_{\mathrm{ac}} = 0.443$~meV. We choose $\Delta_{\mathrm{xx}} = -2.5$~meV, and $\hbar A_{\mathrm{L}} = 1.5$~meV, which do not fully correspond to the parameters studied for the two-level system, but the frequency of the acoustic field remains similar to that in the previous section. In this case, the acoustic field amplitude matches the one we needed to prepare the exciton state, but the energy scale differs. The absence of resonant optical coupling between the ground and biexciton states and coupling of both these states to $\ket{\mathrm{x}}$ results in lower splitting of the corresponding dressed states and thus lower Rabi frequency.
The obtained evolution of state occupations is shown in \figref{fig:biexcitonMap}(b) (solid lines), again together with the pulse envelopes (dashed lines), where a perfect swing-up preparation of the biexciton state is achieved with only a transient occupation of the exciton state during the optical pulse.

\subsection*{Phonon-induced decoherence}\label{sec:decoherence}

The evolution of a real system differs from the idealized one studied above. First, there is the obvious impact of the radiative recombination from the exciton and biexciton states, which forces one to prepare the state in a period much shorter than the recombination time. This limits the maximal post-pulse occupation produced by the protocol, for which we account by solving Eq.~\eqref{eq:Lindblad}.
A more subtle effect arises due to the presence of a phonon environment. As both exciton and biexciton states couple to deformations, phonons can dynamically react to the evolution of charge states in a QD \cite{Roszak2005}. This reaction creates system-bath entanglement and, consequently, leads to the decoherence of QD states. This process is non-Markovian and requires careful treatment. Calculations in the 4th-order correlation expansion \cite{Bracht2022}, path-integral method \cite{Vagov2011}, and tensor-network approaches \cite{Cygorek2022, Strathearn2018} showed that this decoherence can be particularly strong when the relevant evolution frequency falls in the range of strong phonon response. In our study, we maintain the acoustic frequency within the range of weak to moderate phonon response, which enables us to safely work in the second-order Born approximation. We estimate the fidelity $F$ of the prepared states by calculating the post-protocol difference in density matrices between unperturbed evolution and one that includes phonon response. For this, we follow the approach from Ref.~\cite{Roszak2005}, which we outline in the Methods section with more standard derivations provided in Supplementary Note 2~\cite{supplement}.

To relate our study to existing and application-relevant platforms, we consider two types of semiconductor QDs: standard self-assembled InAs/GaAs QDs and GaAs/Al$_{0.4}$Ga$_{0.6}$As QDs grown by droplet etching epitaxy \cite{daSilva2021}. Details of the calculation and QD modeling are given in the Methods section. We calculate the post-pulse occupation of the desired state and fidelity for the cases of exciton and biexciton preparation with Gaussian optical pulses during continuous-wave acoustic modulation, which is the case more prone to non-idealities due to temporal overlaps between acoustic modulation and the (un)dressing of states with light.

\begin{figure}
    \centering
    \includegraphics[width=0.5\columnwidth]{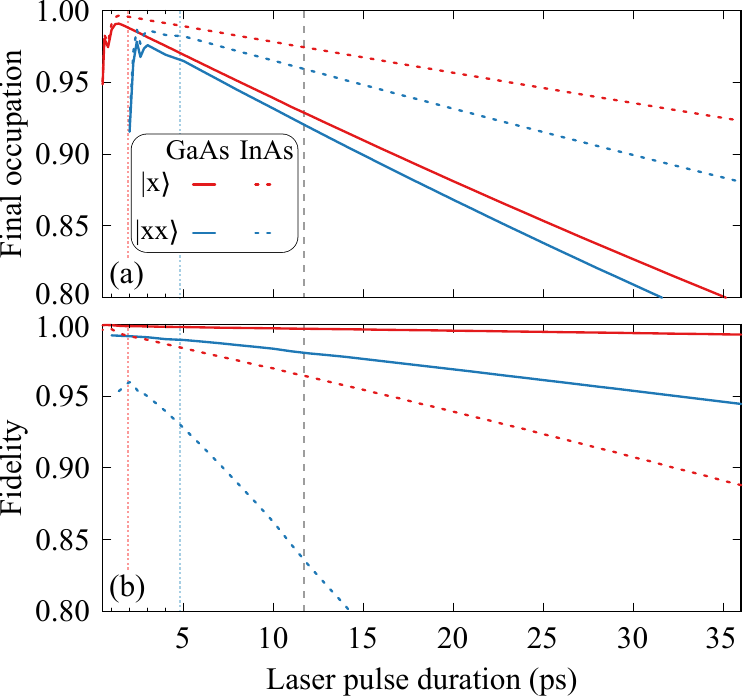}
    \caption{\textbf{Impact of laser pulse duration.} (a) Maximal post-pulse occupation of the desired state and (b) the fidelity of this final state calculated as a function of the laser pulse duration $\sigma_{\mathrm{L}}$. Solid (dashed) lines correspond to GaAs (InAs) QDs; red (blue) lines are for the exciton (biexciton) state. Recombination times used: $\tau_{\mathrm{x}}^{\mathrm{(GaAs)}}=0.426$~ns, $\tau_{\mathrm{xx}}^{\mathrm{(GaAs)}}=0.39$~ns \cite{Heyn2012}, $\tau_{\mathrm{x}}^{\mathrm{(InAs)}}=1.22$~ns, $\tau_{\mathrm{xx}}^{\mathrm{(InAs)}}=0.76$~ns \cite{Feucker2008}. The vertical dashed line marks $\sigma_{\mathrm{L}}=11.7$~ps, while the pale red (blue) vertical dotted line marks the optimal value for the $\ket{\text{x}}$ ($\ket{\text{xx}}$) state.}
    \label{fig:fidelity_finalOccupation}
\end{figure}

First, in \subfigref{fig:fidelity_finalOccupation}{a}, we show the post-pulse occupation of the prepared states as a function of the pulse duration found by numerically solving \eqnref{eq:Lindblad} that accounts for recombination. As expected, shortening the pulse is generally favorable. However, it turns out that it can be shortened even below a few periods of acoustic modulation, which we had previously assumed to be a safe limit, down to the point of single acoustic cycles, below which the method no longer works. We mark $\sigma_{\mathrm{L}}=11.7$~ps, which is maintained throughout most of the discussion in the paper, with a vertical dashed line, while dotted lines indicate the values we found to be optimal. It can be noticed that, characterized by slower recombination, InAs QDs enable the use of longer pulses while still achieving high occupations. Next, we analyze the fidelity of the prepared states (with respect to a pure state with a given occupation of the desired state), as shown in \subfigref{fig:fidelity_finalOccupation}{b}, again versus the pulse duration. Here, we observe that much higher fidelities are achievable for GaAs QDs, which we explain further on, with arbitrarily high $\ket{\text{x}}$ state fidelity for short enough pulses.

\begin{figure}[t]
    \centering
    \makebox[\textwidth]{%
    \includegraphics[width=\textwidth]{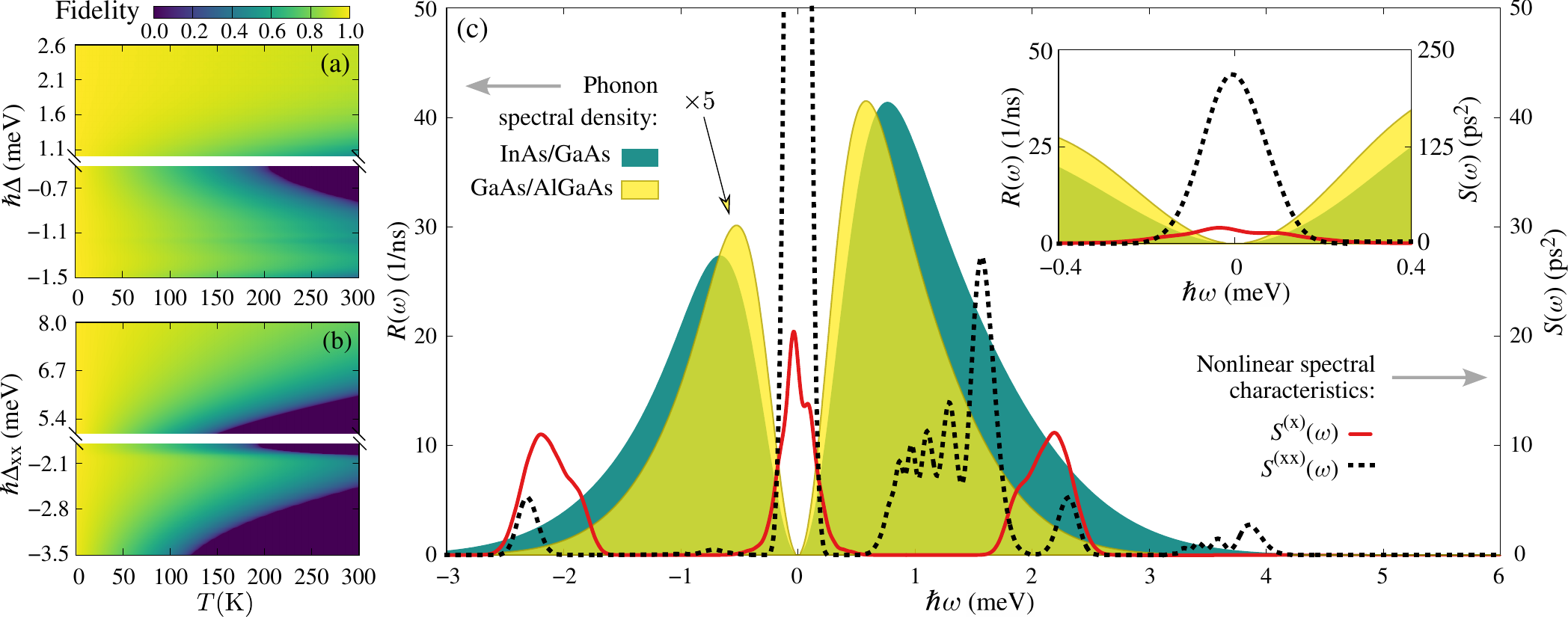}
    }
    \caption{\textbf{State preparation fidelity and decoherence source.} Fidelity of the protocol as a function of the detuning and temperature for GaAs/AlGaAs QDs for (a) exciton preparation, (b) biexciton preparation; (c) Spectral functions calculated for $T = 20$~K. The solid red and dotted black lines are nonlinear spectral characteristics $S(\omega)$ calculated for exciton and biexciton preparation, respectively. Teal- and yellow-shaded areas show phonon spectral densities for InAs/GaAs and GaAs/AlGaAs QDs, respectively. Values for GaAs QDs are low and are scaled $5\times$ for visibility. Note that the central peak in $S(\omega)$ extends beyond the right $y$-axis range, which is chosen narrow to better show the lower side peaks, while the inset repeats the central part of the graph to show those peaks around $\omega=0$.}
    \label{fig:spectralFunctions}
\end{figure}

Due to the predicted significantly better fidelity for GaAs QDs, we explore their case further by calculating fidelity over ranges of temperature and detuning, while keeping the sub-optimal pulse length of $11.7$~ps (see Supplementary Note 3\cite{supplement} for more results for InAs QDs). The results are presented in \subfigref{fig:spectralFunctions}{a} and \subfigref{fig:spectralFunctions}{b} for exciton and biexciton preparation, respectively. In all the cases, we generally observe higher fidelity for higher absolute detuning values. Additionally, the temperature has a crucial impact, with highly preserved fidelity up to $T \sim 100$~K for the exciton and $T \sim 50$~K for the biexciton. However, for specific ranges of detunings, a rapid decrease in fidelity is noticeable. For $\Delta > 0$, this is simply caused by the strong carrier-phonon interaction in the low detuning range, while for $\Delta < 0$, this drop is caused by the mixing of the dressed states, producing similar energy differences between the desired and unwanted transitions. Additionally, the phonon-induced decoherence still has a significant impact here.

The strength of decoherence can be understood based on its physical origin, which stems from the dynamical phonon response to the evolution of charges (polaron formation), thereby entangling the system with its environment. One can decompose this evolution into contributions with different frequencies, which is described by the nonlinear spectral function $S(\omega)$ (see Supplementary Note 2\cite{supplement} for a formal definition). Phonons can respond only at a limited range of frequencies, and their frequency-dependent response strength is described with spectral density $R(\omega)$ (see Supplementary Note 2\cite{supplement} again), which also depends on the geometry of confined charge states. The loss of fidelity is given by the overlap between those two spectral functions. We present them in \subfigref{fig:spectralFunctions}{c}. The spectral densities for the two kinds of QDs are plotted with filled curves for $T = 20$~K, and  $S(\omega)$ is shown for both exciton (solid red line) and biexciton (dotted black line) preparation. One may notice multiple peaks over a wide range of frequencies (energy), with a much richer $S(\omega)$ spectrum for the biexciton preparation protocol. The multiple peaks in the range from $0.75$~meV to $2$~meV correspond to the transitions between exciton and biexciton states during the evolution. Phonon spectral densities strongly depend on the QD geometry, and $R(\omega)$ is much wider and higher for small InAs/GaAs QDs compared to the large-volume GaAs/AlGaAs QDs. Thus, it covers a larger set of frequencies present in the system evolution, lowering the value of the fidelity. The broadening of $S(\omega)$ peaks around $\omega = 0$ comes from the finite duration of the laser pulse, while the rest of the visible peaks can be assigned to the characteristic frequencies of the unitary evolution of the states and the acoustic field frequency. In particular, the peaks around $\hbar\omega \approx \pm 2$~meV correspond to the optical Rabi frequency between dressed states, to which we tune the acoustic field frequency. The peak around $\hbar\omega \approx 4$~meV agrees with the detuning of the exciton state during the preparation of the biexciton.

Note that by properly choosing the detuning, we avoid overlapping the main evolution frequencies with the maxima of $R(\omega)$, which would lead to significant dephasing that we have checked to be reproduced even in our 2nd-order formalism (not shown), in agreement with less approximate methods applied to the all-optical case \cite{Bracht2022, Vagov2011, Cygorek2022, Strathearn2018}. Thus, one way to even further increase fidelity may be to blueshift the evolution frequencies more. This requires higher laser detuning and, therefore, higher acoustic frequencies, which emphasizes the importance of developing THz acoustic technology.

\begingroup
\renewcommand{\arraystretch}{1.1}
\begin{table}[tb!]
\makebox[\textwidth]{%
\begin{minipage}{0.475\textwidth}\small
    \begin{tabular}{c cc cc}
    \toprule
        & \multicolumn{4}{c} {Proposed acousto-optical} \\ \hline\\[-8pt]
           \multirow{2}*{$T$ (K) }& \multicolumn{2}{c }{~$\ket{\mathrm{x}}$ fidelity (\%)} &  \multicolumn{2}{c}{~$\ket{\mathrm{xx}}$ fidelity (\%)}  \\
            & GaAs & InAs & GaAs & InAs \\ \hline\\[-8pt]
         1 & 99.91 & 99.35 & 99.18 & 93.93 \\
         4  & 99.88 & 99.20 & 98.94 & 93.03 \\
         10 & 99.76 & 98.67 & 98.23 & 89.53 \\
         20 & 99.53 & 97.57 & 96.90 & 82.16 \\
         77 & 98.22 & 90.66 & 88.69 &       \\
         \color{gray}{300} & \color{gray}{92.86} & \color{gray}{55.50} & \color{gray}{43.36} & \color{gray}{}\\\botrule
    \end{tabular}
\end{minipage}
\begin{minipage}{0.02\textwidth}\small
      \centering
        \rule{0.4pt}{13em}
\end{minipage}
\begin{minipage}{0.4\textwidth}\small

    \begin{tabular}{cc cc}
    \toprule
            \multicolumn{4}{c} {Resonant/two-photon} \\ \hline\\[-8pt]
           \multicolumn{2}{c }{~$\ket{\mathrm{x}}$ fidelity (\%)} &  \multicolumn{2}{c}{~$\ket{\mathrm{xx}}$ fidelity (\%)}  \\
        GaAs & InAs & GaAs & InAs \\ \hline\\[-8pt]
         99.92 & 99.71 & 99.78 & 98.46 \\
         99.86 & 99.47 & 99.52 & 97.64 \\
         99.68 & 98.81 & 98.93 & 95.65 \\
         99.36 & 97.67 & 97.91 & 91.99 \\
         97.55 & 90.79 & 91.82 & 66.72 \\
         \color{gray}{90.08} & \color{gray}{56.28} & \color{gray}{62.49} & \color{gray}{}\\\botrule
         
    \end{tabular}
\end{minipage}
}
    \caption{\justifying \textbf{Fidelity of prepared states.} Left part of the table shows the fidelity of the exciton $\ket{\text{x}}$ and biexciton $\ket{\text{xx}}$ states prepared with our method using Gaussian optical pulses and continuous-wave acoustic modulation, calculated for GaAs/AlGaAs and InAs/GaAs QDs at different temperatures. All calculations are done within the three-level model for optimal laser pulse durations of $\sigma_{\mathrm{L}}=1.9$~ps and 4.8~ps for the $\ket{\text{x}}$ and $\ket{\text{xx}}$ states, respectively. For comparison, the right part of the table shows results obtained for resonant (exciton) and two-photon (biexciton) preparation schemes simulated with the same optical pulse duration as used for our method.}
    \label{tab:fidelity}
\end{table}
\endgroup

Finally, we leverage the observation made when analyzing \figref{fig:fidelity_finalOccupation}, and use the uncovered optimal laser pulse durations of $\sigma_{\mathrm{L}}=1.9$~ps and 4.8~ps for preparing the exciton and biexciton, respectively. The calculated temperature dependence of achievable fidelities is shown in the left part of Table~\ref{tab:fidelity}.
In all the cases, we find a very high quality of the prepared states at low temperatures. For InAs/GaAs QDs, we may notice a deterioration with increasing temperature due to a significant increase in phonon occupations. GaAs/AlGaAs QDs show excellent fidelity values at cryogenic temperatures for both exciton and biexciton preparation, and even fidelity exceeding $92.8\%$ at room temperature for exciton preparation. However, this result only considers the evolution of the three-level system and the resultant phonon impact (upper bound to the fidelity). We thus present the room-temperature results in gray to emphasize their limited reliability, as thermal excitations to higher orbital states play a significant role in the evolution at such elevated temperatures, especially in large-volume GaAs QDs~\cite {Lehner2023}.

The right part of Table~\ref{tab:fidelity} shows the fidelity calculated for the standard optical methods that can be considered as a reference: resonant excitation of the exciton and two-photon excitation of the biexciton. In all cases, we use the same pulse duration as in simulations of our method to ensure direct comparability of the results. It is worth noting that in most cases, our scheme performs comparably or even better than the reference methods, while being strongly non-resonant. Overall, for an appropriately selected detuning, our protocol itself proves to be almost decoherence-free in terms of dynamical phonon-induced effects, in agreement with recent findings for the all-optical implementation \cite{Bracht2022, Bracht2023b}. 

\section*{Discussion}\label{sec:conclusions}
We have proposed a new hybrid acousto-optical method for the preparation of exciton and biexciton states and general charge state control in quantum emitters like semiconductor QDs. The method extends the recently introduced swing-up scheme, for which an all-optical implementation was originally proposed. Here, we propose to use the acoustic field to directly modulate the detuning in resonance with the Rabi frequency under off-resonant optical driving and show that it leads to the desired coherent qubit evolution. Our scheme is suitable for both positive and negative detunings, with the primary limitation being the availability of the required acoustic field frequencies. We thus highlight the potential advantage of developing coherent acoustic fields in the THz range. In agreement with similar evaluations for the all-optical implementation, we have found that the scheme can be almost phonon decoherence-free even at elevated temperatures.

We have shown the method to work in the weak modulation regime, in which the central spectral QD line remains unshifted and retains more than 99\% of the total weight, and only a fraction of the emission is redistributed into sidebands \cite{Wigger2021}. Thus, any impact on, e.g., photon indistinguishability can be mitigated by standard filtering or post-selection techniques, especially given large modulation frequencies.
On the other hand, sufficiently strong modulation could be used to dynamically detune the QD from, e.g., a cavity resonance \cite{Wigger2017}, to temporarily suppress recombination, similarly to what has been exploited in the all-optical swing-up method with strong optical pulses \cite{Heinish2024, Bracht2023b}.

The proposed hybrid scheme offers a novel off-resonant excitation and charge qubit control method for quantum dots, which does not rely on pulse shaping. More importantly, it also enables acoustic state control when using the acoustic field as a trigger during a longer period of optical driving, even though the system lacks direct acoustic coupling between the charge states. The opposite scheme, i.e., the optical pulse during continuous-wave acoustic modulation, in the longer term, opens a path to producing optically controlled entanglement between the emitter and a quantum acoustic mode and further to state transfer via an acoustic bus, laying the foundations for future acoustically coupled multi-component quantum devices.

\section*{Methods}
\subsection*{Phonon-induced decoherence in the three-level model of a QD}\label{sec:app-decoh}

In Supplementary Note 2~\cite{supplement}, we provide standard derivations of all required equations to model the phonon-induced decoherence of a charge qubit. Here, we summarize the most important part and provide the material parameters used for numerical simulations.

The loss of the fidelity for a system coupled to the phonon bath can be written as \cite{Roszak2005, Kawa2022}
\begin{equation}
    F^2 = 1-\int\mathrm{d}\omega S(\omega)R(\omega),
\end{equation}
where $S(\omega)$ and $R(\omega)$ are the nonlinear spectral characteristic and spectral density of the reservoir, respectively. For carrier-phonon interaction, we use the standard independent boson model with the interaction Hamiltonian $V = S\otimes R$, where \cite{Gawarecki2012}
\begin{equation}
    S = \ketbra{\mathrm{x}}{\mathrm{x}} + 2\ketbra{\mathrm{xx}}{\mathrm{xx}},
\end{equation}
and
\begin{equation}
    R = \sum_{\bm{k}, \lambda}F(\bm{k})\left(b_{\bm{k}, \lambda} + b_{-\bm{k}, \lambda}^{\dagger}\right),
\end{equation}
where $F(\bm{k}) = F(-\bm{k})$. Here $b_{\bm{k}, \lambda}$ and $b_{\bm{k}, \lambda}^{\dagger}$ are the annihilation and creation operators for phonons in a mode labeled by $\bm{k}$ and branch (polarization) $\lambda$ with three values, LA, TA1, TA2 corresponding to longitudinal, and two transverse acoustic branches.

\begin{table}[tbp!]
\begin{tabular}{l|ccccccccc}
\toprule
  \multirow{2}*{QDs}& $a_c$ & $a_v$ & $d$ & $c_{\mathrm{LA}}$ & $e_{14}$ & $\varepsilon_r$ & $l_\rho$  & $l_z$ \\[-3pt]
   & (eV) & (eV) & (kg/$\mathrm{m}^3$) & (m/s) & (C/$\mathrm{m}^2$) & &  (nm) & (nm) \\ \hline \\[-8pt]
InAs & $-7.17$ & $-1.16$ & 5350 & 4730 & $-0.16$ & 12.9 & 6 & 2  \\
GaAs & $-6.56$ & $-0.29$ & 4698 & 5530 & $-0.145$ & 11.4 & 9 & 3  \\
\botrule
\end{tabular}
    \caption{\textbf{Parameters used in calculations.} Material parameters from Refs.~\cite{Tse2013, Levinshtein1999, Vurgaftman2001} and structural parameters used to calculate phonon spectral densities. For the AlGaAs barrier, linear interpolation between AlAs and GaAs is used. Material parameters refer to the barrier materials, as we consider bulk phonons.}
    \label{tab:material}
\end{table}

The corresponding spectral density $R(\omega)$ is given by
\begin{equation}
    R(\omega) = \frac{\mathcal{V}|n(\omega) + 1|}{(2\pi)^3 \hbar^2}\sum_{\lambda}\int_{\mathbb{R}^3}\mathrm{d}^3\bm{k}\left|g_{\bm{k, \lambda}}\right|^2\delta\left(|\omega| - \omega_{\bm{k}, \lambda}\right),
\end{equation}
where $\mathcal{V}$ is the phonon mode normalization volume, $n(\omega)$ is the Bose-Einstein distribution, and the coupling $g_{\bm{k, \lambda}}$ between phonons and an electron-hole pair
\begin{equation}
    g_{\bm{k, \lambda}} = g_{\bm{k, \lambda}}^{(\mathrm{e, DP})} + g_{\bm{k, \lambda}}^{(\mathrm{e, PE})} + g_{\bm{k, \lambda}}^{(\mathrm{h, DP})} + g_{\bm{k, \lambda}}^{(\mathrm{h, PE})}
\end{equation}
contains couplings by piezoelectric effect (PE) and deformation potential (DP). We also assume linear phonon dispersion $\omega_{\bm{k}, \lambda} = c_{\lambda}k$, where $c_{\lambda}$ is the sound velocity in the material. The coupling constants can be written as
\begin{subequations}
    \begin{equation}
        g_{\bm{k, \lambda}}^{(\mathrm{e/h, DP})} = a_{c/v}\sqrt{\frac{\hbar k}{2d\mathcal{V}c_{\mathrm{\lambda}}}}\mathcal{F}_{\mathrm{e/h}}(\bm{k})\delta_{\lambda, \mathrm{LA}},
    \end{equation}
    \begin{equation}
        g_{\bm{k, \lambda}}^{(\mathrm{e/h, PE})} = \pm i\frac{ee_{14}}{\varepsilon_{0}\varepsilon_{r}}\sqrt{\frac{\hbar}{2d\mathcal{V}\omega_{\bm{k},\lambda}}}M_{\lambda}(\hat{\bm{k}}) \mathcal{F}_{\mathrm{e/h}}(\bm{k}),
    \end{equation}
\end{subequations}
where
\begin{equation}
    \mathcal{F}_{\mathrm{e/h}}(\bm{k}) = \int_{\mathbb{R}^3}\mathrm{d}^3\bm{r}\,\varPsi^{*}_{\mathrm{e/h}}(\bm{r})\varPsi_{\mathrm{e/h}}(\bm{r})e^{i\bm{k}\cdot\bm{r}}
\end{equation}
is the form factor that contains all information about the system via the electron/hole ground-state wave functions $\varPsi_{\mathrm{e/h}}(\bm{r})$, and $M_{\lambda}(\hat{\bm{k}})$ is a function that depends on the orientation of the phonon wave vector \cite{Mahan2000},
\begin{equation}
    M_{\lambda}(\hat{\bm{k}}) = 2\hat{k}_x\hat{k}_y\left(\hat{\bm{e}}_{\bm{k}, \lambda}\right)_{z} + \mathrm{c.p.}
\end{equation}
where $\hat{\bm{e}}_{\bm{k}, \lambda}$ is the unit polarization vector for the phonon wave vector $\bm{k}$ with polarization $\lambda$, and $\hat{\bm{k}} = \bm{k}/k$.

For a QD, we assume standard Gaussian wave functions 
\begin{equation}
\varPsi_{\mathrm{e/h}}(\bm{r}) = (2\pi)^{-\frac{3}{2}}\frac{1}{l_z l_\rho^2}\exp\left({-\frac{z^2}{2l_z^2}-\frac{\rho^2}{2l_\rho^2}}\right),
\end{equation}
where $\rho=\sqrt{x^2+y^2}$, and $l_z$ and $l_\rho$ are the out-of-plane and in-plane extensions. We assume stronger hole localization compared to the electron, i.e., $l_i^{(\mathrm{h})} = 0.8 l_i^{(\mathrm{e})}$. The material parameters used to calculate phonon spectral densities are shown in Table~\ref{tab:material}.

\backmatter

\bmhead{Supplementary information}
For additional information, see Supplementary Information.

\bmhead{Acknowledgements}
M.~K. acknowledges the support by Grant No. 2023/49/N/ST3/03931 from the National Science Centre (Poland).
M.~G. acknowledges the financing of the MEEDGARD project funded within the QuantERA II Program that has received funding from the European Union's Horizon 2020 research and innovation program under Grant Agreement No. 101017733 and the National Centre for Research and Development, Poland -- project No. QUANTERAII/2/56/MEEDGARD/2024. 

\bmhead{Author contribution}
M.\,G. conceived the model with input from P.\,M. All authors contributed to the methodology. M.\,K. carried out analytical and numerical calculations under the guidance of M.\,G. M.\,K. prepared the first draft of the manuscript. M.\,G. and P.\,M. supervised the project.
All authors discussed the results and contributed to the manuscript.

\bmhead{Competing interests}
The authors declare no competing financial or non-financial interests.

\bmhead{Data availability}
No empirical data were generated or analyzed in the presented research.


\begin{thebibliography}{10}
	\expandafter\ifx\csname url\endcsname\relax
	\def\url#1{\burl{#1}}\fi
	\expandafter\ifx\csname urlprefix\endcsname\relax\def\urlprefix{URL }\fi
	\providecommand{\bibinfo}[2]{#2}
	\providecommand{\eprint}[2][]{\url{#2}}
	\providecommand{\doi}[1]{\url{https://doi.org/#1}}
	\bibcommenthead
	
	\bibitem{Alfieri2022}
	\bibinfo{author}{Alfieri, A.}, \bibinfo{author}{Anantharaman, S.~B.},
	\bibinfo{author}{Zhang, H.} \& \bibinfo{author}{Jariwala, D.}
	\newblock \bibinfo{title}{Nanomaterials for quantum information science and
		engineering}.
	\newblock \emph{\bibinfo{journal}{Adv. Mater.}} \textbf{\bibinfo{volume}{35}}
	(\bibinfo{year}{2022}).
	
	\bibitem{Roth2023}
	\bibinfo{author}{Roth, N.} \& \bibinfo{author}{Goodwin, A.~L.}
	\newblock \bibinfo{title}{Tuning electronic and phononic states with hidden
		order in disordered crystals}.
	\newblock \emph{\bibinfo{journal}{Nat. Commun.}} \textbf{\bibinfo{volume}{14}},
	\bibinfo{pages}{4328} (\bibinfo{year}{2023}).
	
	\bibitem{Preuss2022}
	\bibinfo{author}{Preuss, J.~A.} \emph{et~al.}
	\newblock \bibinfo{title}{Resonant and phonon-assisted ultrafast coherent
		control of a single h{BN} color center}.
	\newblock \emph{\bibinfo{journal}{Optica}} \textbf{\bibinfo{volume}{9}},
	\bibinfo{pages}{522--531} (\bibinfo{year}{2022}).
	
	\bibitem{Choquer2022}
	\bibinfo{author}{Choquer, M.} \emph{et~al.}
	\newblock \bibinfo{title}{Quantum control of optically active artificial atoms
		with surface acoustic waves}.
	\newblock \emph{\bibinfo{journal}{IEEE Trans. Quantum Eng.}}
	\textbf{\bibinfo{volume}{3}}, \bibinfo{pages}{1--17} (\bibinfo{year}{2022}).
	
	\bibitem{Wang2018}
	\bibinfo{author}{Wang, L.} \emph{et~al.}
	\newblock \bibinfo{title}{{High performance 33.7 GHz surface acoustic wave
			nanotransducers based on AlScN/diamond/Si layered structures}}.
	\newblock \emph{\bibinfo{journal}{Appl. Phys. Lett.}}
	\textbf{\bibinfo{volume}{113}}, \bibinfo{pages}{093503}
	(\bibinfo{year}{2018}).
	
	\bibitem{Maznev2012}
	\bibinfo{author}{Maznev, A.} \emph{et~al.}
	\newblock \bibinfo{title}{Broadband terahertz ultrasonic transducer based on a
		laser-driven piezoelectric semiconductor superlattice}.
	\newblock \emph{\bibinfo{journal}{Ultrasonics}} \textbf{\bibinfo{volume}{52}},
	\bibinfo{pages}{1--4} (\bibinfo{year}{2012}).
	
	\bibitem{Barajas-Aguilar2024}
	\bibinfo{author}{Barajas-Aguilar, A.~H.} \emph{et~al.}
	\newblock \bibinfo{title}{Electrically driven amplification of terahertz
		acoustic waves in graphene}.
	\newblock \emph{\bibinfo{journal}{Nat. Commun.}} \textbf{\bibinfo{volume}{15}},
	\bibinfo{pages}{2550} (\bibinfo{year}{2024}).
	
	\bibitem{Akimov2015}
	\bibinfo{author}{Akimov, A.}, \bibinfo{author}{Scherbakov, A.},
	\bibinfo{author}{Yakovlev, D.} \& \bibinfo{author}{Bayer, M.}
	\newblock \bibinfo{title}{Picosecond acoustics in semiconductor optoelectronic
		nanostructures}.
	\newblock \emph{\bibinfo{journal}{Ultrasonics}} \textbf{\bibinfo{volume}{56}},
	\bibinfo{pages}{122--128} (\bibinfo{year}{2015}).
	
	\bibitem{Ruello2015}
	\bibinfo{author}{Ruello, P.} \& \bibinfo{author}{Gusev, V.~E.}
	\newblock \bibinfo{title}{Physical mechanisms of coherent acoustic phonons
		generation by ultrafast laser action}.
	\newblock \emph{\bibinfo{journal}{Ultrasonics}} \textbf{\bibinfo{volume}{56}},
	\bibinfo{pages}{21--35} (\bibinfo{year}{2015}).
	
	\bibitem{Delsing2019}
	\bibinfo{author}{Delsing, P.} \emph{et~al.}
	\newblock \bibinfo{title}{The 2019 surface acoustic waves roadmap}.
	\newblock \emph{\bibinfo{journal}{J. Phys. D: Appl. Phys.}}
	\textbf{\bibinfo{volume}{52}}, \bibinfo{pages}{353001}
	(\bibinfo{year}{2019}).
	
	\bibitem{Clark2024}
	\bibinfo{author}{Clark, G.} \emph{et~al.}
	\newblock \bibinfo{title}{Nanoelectromechanical control of spin–photon
		interfaces in a hybrid quantum system on chip}.
	\newblock \emph{\bibinfo{journal}{Nano Lett.}} \textbf{\bibinfo{volume}{24}},
	\bibinfo{pages}{1316--1323} (\bibinfo{year}{2024}).
	
	\bibitem{Dietz2023}
	\bibinfo{author}{Dietz, J.~R.}, \bibinfo{author}{Jiang, B.},
	\bibinfo{author}{Day, A.~M.}, \bibinfo{author}{Bhave, S.~A.} \&
	\bibinfo{author}{Hu, E.~L.}
	\newblock \bibinfo{title}{Spin-acoustic control of silicon vacancies in 4{H}
		silicon carbide}.
	\newblock \emph{\bibinfo{journal}{Nat. Electron.}}
	\textbf{\bibinfo{volume}{6}}, \bibinfo{pages}{739--745}
	(\bibinfo{year}{2023}).
	
	\bibitem{Jadot2021}
	\bibinfo{author}{Jadot, B.} \emph{et~al.}
	\newblock \bibinfo{title}{Distant spin entanglement via fast and coherent
		electron shuttling}.
	\newblock \emph{\bibinfo{journal}{Nat. Nanotechnol.}}
	\textbf{\bibinfo{volume}{16}}, \bibinfo{pages}{570--575}
	(\bibinfo{year}{2021}).
	
	\bibitem{Moores2018}
	\bibinfo{author}{Moores, B.~A.}, \bibinfo{author}{Sletten, L.~R.},
	\bibinfo{author}{Viennot, J.~J.} \& \bibinfo{author}{Lehnert, K.~W.}
	\newblock \bibinfo{title}{Cavity quantum acoustic device in the multimode
		strong coupling regime}.
	\newblock \emph{\bibinfo{journal}{Phys. Rev. Lett.}}
	\textbf{\bibinfo{volume}{120}}, \bibinfo{pages}{227701}
	(\bibinfo{year}{2018}).
	
	\bibitem{Sletten2019}
	\bibinfo{author}{Sletten, L.~R.}, \bibinfo{author}{Moores, B.~A.},
	\bibinfo{author}{Viennot, J.~J.} \& \bibinfo{author}{Lehnert, K.~W.}
	\newblock \bibinfo{title}{Resolving phonon {F}ock states in a multimode cavity
		with a double-slit qubit}.
	\newblock \emph{\bibinfo{journal}{Phys. Rev. X}} \textbf{\bibinfo{volume}{9}},
	\bibinfo{pages}{021056} (\bibinfo{year}{2019}).
	
	\bibitem{Lazić2019}
	\bibinfo{author}{Lazi{\'{c}}, S.} \emph{et~al.}
	\newblock \bibinfo{title}{Dynamically tuned non-classical light emission from
		atomic defects in hexagonal boron nitride}.
	\newblock \emph{\bibinfo{journal}{Commun. Phys.}} \textbf{\bibinfo{volume}{2}},
	\bibinfo{pages}{113} (\bibinfo{year}{2019}).
	
	\bibitem{Pezze2021}
	\bibinfo{author}{Pezz{\`e}, L.}
	\newblock \bibinfo{title}{Entanglement-enhanced sensor networks}.
	\newblock \emph{\bibinfo{journal}{Nat. Photonics}}
	\textbf{\bibinfo{volume}{15}}, \bibinfo{pages}{74--76}
	(\bibinfo{year}{2021}).
	
	\bibitem{Gu2023}
	\bibinfo{author}{Gu, S.-S.} \emph{et~al.}
	\newblock \bibinfo{title}{Probing two driven double quantum dots strongly
		coupled to a cavity}.
	\newblock \emph{\bibinfo{journal}{Phys. Rev. Lett.}}
	\textbf{\bibinfo{volume}{130}}, \bibinfo{pages}{233602}
	(\bibinfo{year}{2023}).
	
	\bibitem{Thomas2024}
	\bibinfo{author}{Thomas, S.~E.} \emph{et~al.}
	\newblock \bibinfo{title}{Deterministic storage and retrieval of telecom light
		from a quantum dot single-photon source interfaced with an atomic quantum
		memory}.
	\newblock \emph{\bibinfo{journal}{Sci. Adv.}} \textbf{\bibinfo{volume}{10}},
	\bibinfo{pages}{eadi7346} (\bibinfo{year}{2024}).
	
	\bibitem{Vajner2022}
	\bibinfo{author}{Vajner, D.~A.}, \bibinfo{author}{Rickert, L.},
	\bibinfo{author}{Gao, T.}, \bibinfo{author}{Kaymazlar, K.} \&
	\bibinfo{author}{Heindel, T.}
	\newblock \bibinfo{title}{Quantum communication using semiconductor quantum
		dots}.
	\newblock \emph{\bibinfo{journal}{Adv. Quantum Technol.}}
	\textbf{\bibinfo{volume}{5}}, \bibinfo{pages}{2100116}
	(\bibinfo{year}{2022}).
	
	\bibitem{Michler2017}
	\bibinfo{author}{Michler, P.}
	\newblock \emph{\bibinfo{title}{Quantum dots for quantum information
			technologies}} Vol. \bibinfo{volume}{237} (\bibinfo{publisher}{Springer},
	\bibinfo{year}{2017}).
	
	\bibitem{Benyoucef2023}
	\bibinfo{author}{Benyoucef, M.} \& \bibinfo{author}{Musiał, A.}
	\newblock \emph{\bibinfo{title}{Telecom Wavelengths InP -Based Quantum Dots for
			Quantum Communication}}  (\bibinfo{publisher}{John Wiley \& Sons, Ltd},
	\bibinfo{year}{2023}).
	
	\bibitem{Kim2020}
	\bibinfo{author}{Kim, T.} \emph{et~al.}
	\newblock \bibinfo{title}{Efficient and stable blue quantum dot light-emitting
		diode}.
	\newblock \emph{\bibinfo{journal}{Nature}} \textbf{\bibinfo{volume}{586}},
	\bibinfo{pages}{385--389} (\bibinfo{year}{2020}).
	
	\bibitem{Meng2022}
	\bibinfo{author}{Meng, T.} \emph{et~al.}
	\newblock \bibinfo{title}{Ultrahigh-resolution quantum-dot light-emitting
		diodes}.
	\newblock \emph{\bibinfo{journal}{Nat. Photonics}}
	\textbf{\bibinfo{volume}{16}}, \bibinfo{pages}{297--303}
	(\bibinfo{year}{2022}).
	
	\bibitem{Hafenbrak2007}
	\bibinfo{author}{Hafenbrak, R.} \emph{et~al.}
	\newblock \bibinfo{title}{Triggered polarization-entangled photon pairs from a
		single quantum dot up to 30{K}}.
	\newblock \emph{\bibinfo{journal}{New J. Phys.}} \textbf{\bibinfo{volume}{9}},
	\bibinfo{pages}{315} (\bibinfo{year}{2007}).
	
	\bibitem{Clerk2020}
	\bibinfo{author}{Clerk, A.~A.}, \bibinfo{author}{Lehnert, K.~W.},
	\bibinfo{author}{Bertet, P.}, \bibinfo{author}{Petta, J.~R.} \&
	\bibinfo{author}{Nakamura, Y.}
	\newblock \bibinfo{title}{Hybrid quantum systems with circuit quantum
		electrodynamics}.
	\newblock \emph{\bibinfo{journal}{Nat. Physics}} \textbf{\bibinfo{volume}{16}},
	\bibinfo{pages}{257--267} (\bibinfo{year}{2020}).
	
	\bibitem{Kim2022}
	\bibinfo{author}{Kim, C.~W.}, \bibinfo{author}{Nichol, J.~M.},
	\bibinfo{author}{Jordan, A.~N.} \& \bibinfo{author}{Franco, I.}
	\newblock \bibinfo{title}{Analog quantum simulation of the dynamics of open
		quantum systems with quantum dots and microelectronic circuits}.
	\newblock \emph{\bibinfo{journal}{PRX Quantum}} \textbf{\bibinfo{volume}{3}},
	\bibinfo{pages}{040308} (\bibinfo{year}{2022}).
	
	\bibitem{Weiss2021}
	\bibinfo{author}{Wei{\ss}, M.} \emph{et~al.}
	\newblock \bibinfo{title}{Optomechanical wave mixing by a single quantum dot}.
	\newblock \emph{\bibinfo{journal}{Optica}} \textbf{\bibinfo{volume}{8}},
	\bibinfo{pages}{291--300} (\bibinfo{year}{2021}).
	
	\bibitem{Wigger2021}
	\bibinfo{author}{Wigger, D.} \emph{et~al.}
	\newblock \bibinfo{title}{Resonance-fluorescence spectral dynamics of an
		acoustically modulated quantum dot}.
	\newblock \emph{\bibinfo{journal}{Phys. Rev. Res.}}
	\textbf{\bibinfo{volume}{3}}, \bibinfo{pages}{033197} (\bibinfo{year}{2021}).
	
	\bibitem{Stievater2001}
	\bibinfo{author}{Stievater, T.~H.} \emph{et~al.}
	\newblock \bibinfo{title}{Rabi oscillations of excitons in single quantum
		dots}.
	\newblock \emph{\bibinfo{journal}{Phys. Rev. Lett.}}
	\textbf{\bibinfo{volume}{87}}, \bibinfo{pages}{133603}
	(\bibinfo{year}{2001}).
	
	\bibitem{Kamada2001}
	\bibinfo{author}{Kamada, H.}, \bibinfo{author}{Gotoh, H.},
	\bibinfo{author}{Temmyo, J.}, \bibinfo{author}{Takagahara, T.} \&
	\bibinfo{author}{Ando, H.}
	\newblock \bibinfo{title}{Exciton {R}abi oscillation in a single quantum dot}.
	\newblock \emph{\bibinfo{journal}{Phys. Rev. Lett.}}
	\textbf{\bibinfo{volume}{87}}, \bibinfo{pages}{246401}
	(\bibinfo{year}{2001}).
	
	\bibitem{Thomas2021}
	\bibinfo{author}{Thomas, S.~E.} \emph{et~al.}
	\newblock \bibinfo{title}{Bright polarized single-photon source based on a
		linear dipole}.
	\newblock \emph{\bibinfo{journal}{Phys. Rev. Lett.}}
	\textbf{\bibinfo{volume}{126}}, \bibinfo{pages}{233601}
	(\bibinfo{year}{2021}).
	
	\bibitem{Vyvlecka2023}
	\bibinfo{author}{Vyvlecka, M.} \emph{et~al.}
	\newblock \bibinfo{title}{{Robust excitation of C-band quantum dots for quantum
			communication}}.
	\newblock \emph{\bibinfo{journal}{Appl. Phys. Lett.}}
	\textbf{\bibinfo{volume}{123}}, \bibinfo{pages}{174001}
	(\bibinfo{year}{2023}).
	
	\bibitem{Simon2011}
	\bibinfo{author}{Simon, C.-M.} \emph{et~al.}
	\newblock \bibinfo{title}{Robust quantum dot exciton generation via adiabatic
		passage with frequency-swept optical pulses}.
	\newblock \emph{\bibinfo{journal}{Phys. Rev. Lett.}}
	\textbf{\bibinfo{volume}{106}}, \bibinfo{pages}{166801}
	(\bibinfo{year}{2011}).
	
	\bibitem{Kaldeway2017}
	\bibinfo{author}{Kaldewey, T.} \emph{et~al.}
	\newblock \bibinfo{title}{Demonstrating the decoupling regime of the
		electron-phonon interaction in a quantum dot using chirped optical
		excitation}.
	\newblock \emph{\bibinfo{journal}{Phys. Rev. B}} \textbf{\bibinfo{volume}{95}},
	\bibinfo{pages}{241306} (\bibinfo{year}{2017}).
	
	\bibitem{Stufler2006}
	\bibinfo{author}{Stufler, S.} \emph{et~al.}
	\newblock \bibinfo{title}{Two-photon {R}abi oscillations in a single
		{I}n$_{x}${G}a$_{1-x}\mathrm{As}/\mathrm{Ga}\mathrm{As}$ quantum dot}.
	\newblock \emph{\bibinfo{journal}{Phys. Rev. B}} \textbf{\bibinfo{volume}{73}},
	\bibinfo{pages}{125304} (\bibinfo{year}{2006}).
	
	\bibitem{Machnikowski2008}
	\bibinfo{author}{Machnikowski, P.}
	\newblock \bibinfo{title}{Theory of two-photon processes in quantum dots:
		Coherent evolution and phonon-induced dephasing}.
	\newblock \emph{\bibinfo{journal}{Phys. Rev. B}} \textbf{\bibinfo{volume}{78}},
	\bibinfo{pages}{195320} (\bibinfo{year}{2008}).
	
	\bibitem{He2019}
	\bibinfo{author}{He, Y.-M.} \emph{et~al.}
	\newblock \bibinfo{title}{Coherently driving a single quantum two-level system
		with dichromatic laser pulses}.
	\newblock \emph{\bibinfo{journal}{Nat. Phys.}} \textbf{\bibinfo{volume}{15}},
	\bibinfo{pages}{941--946} (\bibinfo{year}{2019}).
	
	\bibitem{Koong2021}
	\bibinfo{author}{Koong, Z.~X.} \emph{et~al.}
	\newblock \bibinfo{title}{Coherent dynamics in quantum emitters under
		dichromatic excitation}.
	\newblock \emph{\bibinfo{journal}{Phys. Rev. Lett.}}
	\textbf{\bibinfo{volume}{126}}, \bibinfo{pages}{047403}
	(\bibinfo{year}{2021}).
	
	\bibitem{Matthiesen2012}
	\bibinfo{author}{Matthiesen, C.}, \bibinfo{author}{Vamivakas, A.~N.} \&
	\bibinfo{author}{Atat\"ure, M.}
	\newblock \bibinfo{title}{Subnatural linewidth single photons from a quantum
		dot}.
	\newblock \emph{\bibinfo{journal}{Phys. Rev. Lett.}}
	\textbf{\bibinfo{volume}{108}}, \bibinfo{pages}{093602}
	(\bibinfo{year}{2012}).
	
	\bibitem{Ardelt2014}
	\bibinfo{author}{Ardelt, P.-L.} \emph{et~al.}
	\newblock \bibinfo{title}{Dissipative preparation of the exciton and biexciton
		in self-assembled quantum dots on picosecond time scales}.
	\newblock \emph{\bibinfo{journal}{Phys. Rev. B}} \textbf{\bibinfo{volume}{90}},
	\bibinfo{pages}{241404} (\bibinfo{year}{2014}).
	
	\bibitem{Barth2016}
	\bibinfo{author}{Barth, A.~M.} \emph{et~al.}
	\newblock \bibinfo{title}{Fast and selective phonon-assisted state preparation
		of a quantum dot by adiabatic undressing}.
	\newblock \emph{\bibinfo{journal}{Phys. Rev. B}} \textbf{\bibinfo{volume}{94}},
	\bibinfo{pages}{045306} (\bibinfo{year}{2016}).
	
	\bibitem{Bracht2021}
	\bibinfo{author}{Bracht, T.~K.} \emph{et~al.}
	\newblock \bibinfo{title}{Swing-up of quantum emitter population using detuned
		pulses}.
	\newblock \emph{\bibinfo{journal}{PRX Quantum}} \textbf{\bibinfo{volume}{2}},
	\bibinfo{pages}{040354} (\bibinfo{year}{2021}).
	
	\bibitem{Bracht2022}
	\bibinfo{author}{Bracht, T.~K.}, \bibinfo{author}{Seidelmann, T.},
	\bibinfo{author}{Kuhn, T.}, \bibinfo{author}{Axt, V.~M.} \&
	\bibinfo{author}{Reiter, D.~E.}
	\newblock \bibinfo{title}{Phonon wave packet emission during state preparation
		of a semiconductor quantum dot using different schemes}.
	\newblock \emph{\bibinfo{journal}{Phys. Status Solidi B}}
	\textbf{\bibinfo{volume}{259}}, \bibinfo{pages}{2100649}
	(\bibinfo{year}{2022}).
	
	\bibitem{Bracht2023}
	\bibinfo{author}{Bracht, T.~K.} \emph{et~al.}
	\newblock \bibinfo{title}{Dressed-state analysis of two-color excitation
		schemes}.
	\newblock \emph{\bibinfo{journal}{Phys. Rev. B}}
	\textbf{\bibinfo{volume}{107}}, \bibinfo{pages}{035425}
	(\bibinfo{year}{2023}).
	
	\bibitem{Karli2022}
	\bibinfo{author}{Karli, Y.} \emph{et~al.}
	\newblock \bibinfo{title}{Super scheme in action: Experimental demonstration of
		red-detuned excitation of a quantum emitter}.
	\newblock \emph{\bibinfo{journal}{Nano Lett.}} \textbf{\bibinfo{volume}{22}},
	\bibinfo{pages}{6567--6572} (\bibinfo{year}{2022}).
	
	\bibitem{Boos2024}
	\bibinfo{author}{Boos, K.} \emph{et~al.}
	\newblock \bibinfo{title}{Coherent swing-up excitation for semiconductor
		quantum dots}.
	\newblock \emph{\bibinfo{journal}{Adv. Quantum Technol.}}
	\textbf{\bibinfo{volume}{7}}, \bibinfo{pages}{2300359}
	(\bibinfo{year}{2024}).
	
	\bibitem{daSilva2021}
	\bibinfo{author}{da~Silva, S. F.~C.} \emph{et~al.}
	\newblock \bibinfo{title}{Ga{A}s quantum dots grown by droplet etching epitaxy
		as quantum light sources}.
	\newblock \emph{\bibinfo{journal}{Appl. Phys. Lett.}}
	\textbf{\bibinfo{volume}{119}}, \bibinfo{pages}{120502}
	(\bibinfo{year}{2021}).
	
	\bibitem{supplement}
	\bibinfo{note}{See Supplemental Material for additional derivations and details
		of calculations.}
	
	\bibitem{Roszak2005}
	\bibinfo{author}{Roszak, K.}, \bibinfo{author}{Grodecka, A.},
	\bibinfo{author}{Machnikowski, P.} \& \bibinfo{author}{Kuhn, T.}
	\newblock \bibinfo{title}{Phonon-induced decoherence for a quantum-dot spin
		qubit operated by raman passage}.
	\newblock \emph{\bibinfo{journal}{Phys. Rev. B}} \textbf{\bibinfo{volume}{71}},
	\bibinfo{pages}{195333} (\bibinfo{year}{2005}).
	
	\bibitem{Vagov2011}
	\bibinfo{author}{Vagov, A.}, \bibinfo{author}{Croitoru, M.~D.},
	\bibinfo{author}{Gl\"assl, M.}, \bibinfo{author}{Axt, V.~M.} \&
	\bibinfo{author}{Kuhn, T.}
	\newblock \bibinfo{title}{Real-time path integrals for quantum dots: Quantum
		dissipative dynamics with superohmic environment coupling}.
	\newblock \emph{\bibinfo{journal}{Phys. Rev. B}} \textbf{\bibinfo{volume}{83}},
	\bibinfo{pages}{094303} (\bibinfo{year}{2011}).
	
	\bibitem{Cygorek2022}
	\bibinfo{author}{Cygorek, M.} \emph{et~al.}
	\newblock \bibinfo{title}{Simulation of open quantum systems by automated
		compression of arbitrary environments}.
	\newblock \emph{\bibinfo{journal}{Nat. Phys.}} \textbf{\bibinfo{volume}{18}},
	\bibinfo{pages}{662--668} (\bibinfo{year}{2022}).
	
	\bibitem{Strathearn2018}
	\bibinfo{author}{Strathearn, A.}, \bibinfo{author}{Kirton, P.},
	\bibinfo{author}{Kilda, D.}, \bibinfo{author}{Keeling, J.} \&
	\bibinfo{author}{Lovett, B.~W.}
	\newblock \bibinfo{title}{Efficient non-{M}arkovian quantum dynamics using
		time-evolving matrix product operators}.
	\newblock \emph{\bibinfo{journal}{Nat. Commun.}} \textbf{\bibinfo{volume}{9}},
	\bibinfo{pages}{3322} (\bibinfo{year}{2018}).
	
	\bibitem{Heyn2012}
	\bibinfo{author}{Heyn, C.}, \bibinfo{author}{Strelow, C.} \&
	\bibinfo{author}{Hansen, W.}
	\newblock \bibinfo{title}{Excitonic lifetimes in single {G}a{A}s quantum dots
		fabricated by local droplet etching}.
	\newblock \emph{\bibinfo{journal}{New J. Phys.}} \textbf{\bibinfo{volume}{14}},
	\bibinfo{pages}{053004} (\bibinfo{year}{2012}).
	
	\bibitem{Feucker2008}
	\bibinfo{author}{Feucker, M.}, \bibinfo{author}{Seguin, R.},
	\bibinfo{author}{Rodt, S.}, \bibinfo{author}{Hoffmann, A.} \&
	\bibinfo{author}{Bimberg, D.}
	\newblock \bibinfo{title}{Decay dynamics of neutral and charged excitonic
		complexes in single {I}n{A}s/{G}a{A}s quantum dots}.
	\newblock \emph{\bibinfo{journal}{Appl. Phys. Lett.}}
	\textbf{\bibinfo{volume}{92}}, \bibinfo{pages}{063116}
	(\bibinfo{year}{2008}).
	
	\bibitem{Lehner2023}
	\bibinfo{author}{Lehner, B.~U.} \emph{et~al.}
	\newblock \bibinfo{title}{Beyond the four-level model: Dark and hot states in
		quantum dots degrade photonic entanglement}.
	\newblock \emph{\bibinfo{journal}{Nano Lett.}} \textbf{\bibinfo{volume}{23}},
	\bibinfo{pages}{1409--1415} (\bibinfo{year}{2023}).
	
	\bibitem{Bracht2023b}
	\bibinfo{author}{Bracht, T.~K.}, \bibinfo{author}{Cygorek, M.},
	\bibinfo{author}{Seidelmann, T.}, \bibinfo{author}{Axt, V.~M.} \&
	\bibinfo{author}{Reiter, D.~E.}
	\newblock \bibinfo{title}{Temperature-independent almost perfect photon
		entanglement from quantum dots via the super scheme}.
	\newblock \emph{\bibinfo{journal}{Optica Quantum}}
	\textbf{\bibinfo{volume}{1}}, \bibinfo{pages}{103--107}
	(\bibinfo{year}{2023}).
	
	\bibitem{Wigger2017}
	\bibinfo{author}{Wigger, D.}, \bibinfo{author}{Czerniuk, T.},
	\bibinfo{author}{Reiter, D.~E.}, \bibinfo{author}{Bayer, M.} \&
	\bibinfo{author}{Kuhn, T.}
	\newblock \bibinfo{title}{Systematic study of the influence of coherent phonon
		wave packets on the lasing properties of a quantum dot ensemble}.
	\newblock \emph{\bibinfo{journal}{New J. Phys.}} \textbf{\bibinfo{volume}{19}},
	\bibinfo{pages}{073001} (\bibinfo{year}{2017}).
	
	\bibitem{Heinish2024}
	\bibinfo{author}{Heinisch, N.}, \bibinfo{author}{K\"ocher, N.},
	\bibinfo{author}{Bauch, D.} \& \bibinfo{author}{Schumacher, S.}
	\newblock \bibinfo{title}{Swing-up dynamics in quantum emitter cavity systems:
		Near ideal single photons and entangled photon pairs}.
	\newblock \emph{\bibinfo{journal}{Phys. Rev. Res.}}
	\textbf{\bibinfo{volume}{6}}, \bibinfo{pages}{L012017}
	(\bibinfo{year}{2024}).
	
	\bibitem{Kawa2022}
	\bibinfo{author}{Kawa, K.}, \bibinfo{author}{Kuhn, T.} \&
	\bibinfo{author}{Machnikowski, P.}
	\newblock \bibinfo{title}{Coherence limitations in the optical control of the
		singlet-triplet qubit in a quantum dot molecule}.
	\newblock \emph{\bibinfo{journal}{Phys. Rev. B}}
	\textbf{\bibinfo{volume}{106}}, \bibinfo{pages}{125308}
	(\bibinfo{year}{2022}).
	
	\bibitem{Gawarecki2012}
	\bibinfo{author}{Gawarecki, K.} \emph{et~al.}
	\newblock \bibinfo{title}{Dephasing in the adiabatic rapid passage in quantum
		dots: Role of phonon-assisted biexciton generation}.
	\newblock \emph{\bibinfo{journal}{Phys. Rev. B}} \textbf{\bibinfo{volume}{86}},
	\bibinfo{pages}{235301} (\bibinfo{year}{2012}).
	
	\bibitem{Tse2013}
	\bibinfo{author}{Tse, G.} \emph{et~al.}
	\newblock \bibinfo{title}{{Non-linear piezoelectricity in zinc blende GaAs and
			InAs semiconductors}}.
	\newblock \emph{\bibinfo{journal}{J. Appl. Phys.}}
	\textbf{\bibinfo{volume}{114}}, \bibinfo{pages}{073515}
	(\bibinfo{year}{2013}).
	
	\bibitem{Levinshtein1999}
	\bibinfo{author}{Levinshtein, M.}, \bibinfo{author}{Rumyantsev, S.},
	\bibinfo{author}{Shur, M.} \& \bibinfo{author}{Scientific, W.}
	\newblock \emph{\bibinfo{title}{Handbook Series on Semiconductor Parameters:
			Ternary and quaternary III-V compounds}} EBL-Schweitzer
	(\bibinfo{publisher}{World Scientific Publishing Company},
	\bibinfo{year}{1999}).
	
	\bibitem{Vurgaftman2001}
	\bibinfo{author}{{Vurgaftman}, I.}, \bibinfo{author}{{Meyer}, J.~R.} \&
	\bibinfo{author}{{Ram-Mohan}, L.~R.}
	\newblock \bibinfo{title}{{Band parameters for III-V compound semiconductors
			and their alloys}}.
	\newblock \emph{\bibinfo{journal}{J. Appl. Phys.}}
	\textbf{\bibinfo{volume}{89}}, \bibinfo{pages}{5815--5875}
	(\bibinfo{year}{2001}).
	
	\bibitem{Mahan2000}
	\bibinfo{author}{Mahan, G.~D.}
	\newblock \emph{\bibinfo{title}{Many-Particle Physics (Physics of Solids and
			Liquids)}}  (\bibinfo{publisher}{Springer, Berlin}, \bibinfo{year}{2000}).
	
\end{thebibliography}

\end{document}



\title{Hybrid acousto-optical swing-up state control in a quantum dot}

\author{Mateusz Kuniej\,\orcidlink{0000-0001-5476-4856}}
\author{Pawe{\l} Machnikowski\,\orcidlink{0000-0003-0349-1725}}
\author{Micha{\l} Gawe{\l}czyk\,\orcidlink{0000-0003-2299-140X}}
\affiliation{Institute of Theoretical Physics, Wroc\l{}aw University of Science and Technology, 50-370 Wroc\l{}aw, Poland}

\title{Supplementary Information to ``Hybrid acousto-optical swing-up state control in a quantum dot''}

\begin{abstract}
In this Supplementary Information, we provide derivations of the standard rotating wave approximation adopted for our system, describe the details of fidelity calculations, and present additional results on phonon-induced decoherence.
\end{abstract}

\maketitle


\section{S\MakeTextLowercase{upplementary} N\MakeTextLowercase{ote} 1: R\MakeTextLowercase{otating wave approximation in a three-level model of a quantum dot}}\label{sec:app-rwa}
We consider a three-level system with the ground level $\ket{\mathrm{g}}$ and two excited states $\ket{\mathrm{x}}$ and $\ket{\mathrm{xx}}$. Let the $\hbar\omega$ be the transition energy between $\ket{\mathrm{g}}$ and $\ket{\mathrm{x}}$ states, and $\hbar\omega - \hbar\Delta_{\mathrm{B}}$ between the $\ket{\mathrm{x}}$ and $\ket{\mathrm{xx}}$. Then, the Hamiltonian can be written as
\begin{equation}
    H_0 = \hbar\omega\ketbra{\mathrm{x}}{\mathrm{x}} + \hbar(2\omega - \Delta_{\mathrm{B}})\ketbra{\mathrm{xx}}{\mathrm{xx}}.
\end{equation}
Let the QD interact with an external optical field with frequency $\omega_{\mathrm{L}}$. In the dipole approximation, the interaction Hamiltonian can be expressed as 
\begin{equation}
    H_{\mathrm{int}}(t) = -\bm{d}\cdot\bm{E}(t),
\end{equation}
where $\bm{d}$ is the dipole moment operator. Assuming no diagonal coupling between laser and QD states and appropriate linear polarization of the optical field, we can write the full Hamiltonian $H_0 + H_{\mathrm{int}}$ in the form
\begin{equation}
    \begin{split}
        H(t) =&\ \hbar\omega\ketbra{\mathrm{x}}{\mathrm{x}} + \hbar\left(2\omega - \Delta_{\mathrm{B}}\right)\ketbra{\mathrm{xx}}{\mathrm{xx}} \\ &+ \hbar\Omega_{\mathrm{L}}(t)\left(\ketbra{\mathrm{x}}{\mathrm{g}} + \ketbra{\mathrm{x}}{\mathrm{xx}} + \hc\right),
    \end{split}
\end{equation}
where $\Omega_{\mathrm{L}}(t) = \Omega_{\mathrm{L}}^{(0)}(t)\cos(\omega_{\mathrm{L}}t)$ is a classical optical field. Next, we perform a unitary transformation to a rotating frame with $U(t) = e^{S(t)}$  given by
\begin{equation}
    S(t) = i\omega_{\mathrm{L}} t\ketbra{\mathrm{x}}{\mathrm{x}} + 2i\omega_{\mathrm{L}} t\ketbra{\mathrm{xx}}{\mathrm{xx}}.
\end{equation}
The transformed Hamiltonian is
\begin{equation}
    \widetilde{H}(t) = U(t)H(t)U^{\dagger}(t) + i\hbar\Dot{U}(t)U^{\dagger}(t),
\end{equation}
which gives 
\begin{equation}
    \begin{split}
        \widetilde{H}(t) =&\ \hbar\left(\omega - \omega_{\mathrm{L}}\right)\ketbra{\mathrm{x}}{\mathrm{x}} + \hbar\left(2\omega -2\omega_{\mathrm{L}} - \Delta_{\mathrm{B}}\right)\ketbra{\mathrm{xx}}{\mathrm{xx}} \\ &+ \frac{1}{2}\hbar\Omega_{\mathrm{L}}^{(0)}(t)\left(e^{i\omega_{\mathrm{L}}t} + e^{-i\omega_{\mathrm{L}}t}\right) \\ &\times \left(e^{i\omega_{\mathrm{L}}t}\ketbra{\mathrm{x}}{\mathrm{g}} + e^{-i\omega_{\mathrm{L}}t}\ketbra{\mathrm{x}}{\mathrm{xx}} + \hc\right).
    \end{split}
\end{equation}
In the rotating wave approximation (RWA), the terms $\propto \exp(\pm 2i\omega_{\mathrm{L}}t)$ are neglected due to fast oscillating behavior. The difference $\omega_{\mathrm{L}} - \omega = \Delta$ is the detuning of the optical field from the $\ket{\mathrm{g}}$-$\ket{\mathrm{x}}$ transition, and $2\omega_{\mathrm{L}}-(2\omega - \Delta_{\mathrm{B}}) = 2\Delta +  \Delta_{\mathrm{B}}$ is the detuning from the two-photon $\ket{\mathrm{g}}$-$\ket{\mathrm{xx}}$ transition. Finally, the Hamiltonian in the dipole and RWA approximations reads
\begin{equation}
    \begin{split}
        \widetilde{H}(t) =&\ -\hbar\Delta\ketbra{\mathrm{x}}{\mathrm{x}} - \hbar(2\Delta + \Delta_{\mathrm{B}})\ketbra{\mathrm{xx}}{\mathrm{xx}} \\ &+ \frac{1}{2}\hbar\Omega_{\mathrm{L}}^{(0)}(t)(\ketbra{\mathrm{x}}{\mathrm{g}} + \ketbra{\mathrm{x}}{\mathrm{xx}} + \hc).
    \end{split}
\end{equation}

\section{S\MakeTextLowercase{upplementary} N\MakeTextLowercase{ote} 2: F\MakeTextLowercase{idelity calculation}}\label{sec:app-fidelity}
Here, to find the fidelity of the prepared state, we follow the approach from Supplementary Reference~\cite{Roszak2005}. Let the isolated carrier subsystem be described by the Hamiltonian $H_{\mathrm{S}} = \sum_i E_i \ketbra{i}{i}$.
In the non-perturbed case, the unitary evolution of the entire system composed of the studied subsystem and its phonon bath is given by 
\begin{equation}
    U_0(t) = U(t)\otimes e^{-iH_{\mathrm{ph}}t/\hbar},
\end{equation}
where $U(t)= \exp({-iH_{\mathrm{S}}t/\hbar})$ is the unitary evolution operator of the carrier subsystem in the absence of the carrier-phonon interaction and $H_{\mathrm{ph}}$ is the free phonon Hamiltonian. The interaction between carriers and phonons can always be written in the general form
\begin{equation}
    V = \sum_{nm} S_{nm}\otimes R_{nm},
\end{equation}
where $S_{nm} = \ketbra{n}{m}$ acts in the Hilbert space of the carrier subsystem and
\begin{equation}
    R_{nm} = \sum_{\bm{k}, \lambda}F_{nm}(\bm{k})\left(b_{\bm{k}, \lambda} + b_{-\bm{k}, \lambda}^{\dagger}\right)
\end{equation}
acts only on the phonon environment with the coupling functions $F_{nm}(\bm{k}) = F_{mn}(-\bm{k})$. Here $b_{\bm{k}, \lambda}$ and $b_{\bm{k}, \lambda}^{\dagger}$ are the annihilation and creation operators for phonons in a mode labeled by $\bm{k}$ and branch (polarization) $\lambda$ with three values, LA, TA1, TA2, corresponding to longitudinal and two transverse acoustic branches.

We assume that the system is initially in a product state
\begin{equation}
    \varrho(t_0) = \rho_0 \otimes \rho_{\mathrm{R}},
\end{equation}
where $\varrho$ is the full density matrix, $\rho_{0} = \ketbra{\psi_0}{\psi_0}$ is the initial density matrix of the carrier system, and $\rho_{\mathrm{R}}$ is the phonon density matrix in thermal equilibrium.

The evolution of the full density matrix  $\varrho(t)$ is given by the Liouville-von Neumann equation
\begin{equation}
    \Dot{\varrho}(t) = -\frac{i}{\hbar}\left[H_{\mathrm{S}}+ H_{\mathrm{ph}} + V, \varrho(t)\right].
\end{equation}
We are interested in the evolution of the reduced density matrix of the carrier subsystem, $\rho(t)=\mathrm{Tr}_{\mathrm{R}}\varrho(t)$, where $\mathrm{Tr}_{\mathrm{R}}$ is the trace over the environmental degrees of freedom.
We assume that the influence of the environment is weak so that the reduced density matrix has the form
\begin{equation}
    \rho(t) = \rho^{(0)}(t) + \Delta\rho(t),
\end{equation}
where $\Delta\rho(t)$ is the phonon-induced correction to the unperturbed evolution $\rho^{(0)}(t) = U_0(t)\rho_0U_0^{\dagger}(t)$. To quantify the quality of preparation of the given carrier state, we use the fidelity, which measures the overlap between final ($t\to\infty$) states with and without perturbation. It is given by the scalar product of the unperturbed and perturbed density matrices,
\begin{equation}
    \begin{split}
        F^2 = \Tr\rho^{(0)}(\infty)\rho(\infty) &= \bra{\psi_0}U_0^{\dagger}(\infty)\rho(\infty)U_0(\infty)\ket{\psi_0} \\ &= 1 + \bra{\psi_0}\Delta\Tilde{\rho}(\infty)\ket{\psi_0},
    \end{split}
\end{equation}
where $\Delta\Tilde{\rho}(\infty)$ is the perturbation in the interaction picture at the end of the protocol, and it is calculated in the second-order Born approximation,
\begin{equation}
    \Delta\Tilde{\rho}(t) = -\frac{1}{\hbar^2}\int_{t_0}^{t}\mathrm{d}\tau\int_{t_0}^{\tau}\mathrm{d}\tau'\mathrm{Tr}_{\mathrm{R}}\left[\Tilde{V}(\tau), \left[\Tilde{V}(\tau'), \varrho(t_0)\right]\right],
\end{equation}
where $\Tilde{V}(\tau)$ is $V$ written in the interaction picture. At this stage, it is convenient to introduce two types of spectral functions. The first is the spectral density of the reservoir
\begin{equation}
    R_{ijkl}(\omega) = \frac{1}{2\pi}\int_{-\infty}^{\infty}\mathrm{d}t\left\langle R_{ij}(t)R_{kl}\right\rangle e^{i\omega t},
\end{equation}
where the operator $R_{ij}(t)$ is transformed into the interaction picture. The second spectral function is the nonlinear spectral characteristic, defined as
\begin{equation}
    S_{ijkl}(\omega) = \sum_{n}\bra{\psi_0}Y^{\dagger}_{ij}(\omega)\ketbra{\psi_n}{\psi_n}Y_{kl}(\omega)\ket{\psi_0},
\end{equation}
where the set of vectors $\ket{\psi_n}$ spans the space orthogonal to the initial state. The functions $Y_{ij}(\omega)$ are defined as
\begin{equation}
    Y_{ij}(\omega) = \int_{t_0}^{t}\mathrm{d}\tau S_{ij}(\tau)e^{-i\omega \tau},
\end{equation}
where $S_{ij}(t) = U_0^{\dagger}(t)S_{ij}U_0(t)$. Using the above equations, the fidelity can be written in the form
\begin{equation}
    F^2 = 1 - \sum_{ijkl}\int_{-\infty}^{\infty}\mathrm{d}\omega R_{ijkl}(\omega)S_{ijkl}(\omega).
    \label{eq:fidelity}
\end{equation}

\section{S\MakeTextLowercase{upplementary} N\MakeTextLowercase{ote} 3: P\MakeTextLowercase{honon induced decoherence in quantum dots}}

\begin{figure}[tb!]
    \centering
    \includegraphics[width=0.95\columnwidth]{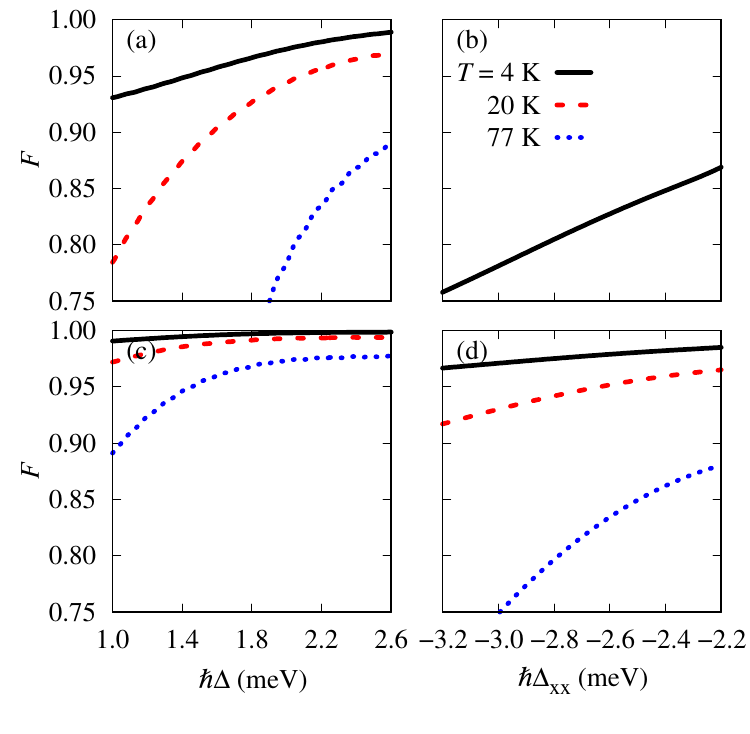}
    \caption{Achievable fidelity as a function of the laser detuning. The top row corresponds to InAs/GaAs QDs, and the bottom row corresponds to GaAs/AlGaAs QDs. The left column stands for the exciton preparation, and the right column for the biexciton case.}
    \label{fig:fidelity}
\end{figure}

We can study the dependence of achievable fidelity on laser detuning with the results for three selected temperatures shown in Supplementary Figure~\ref{fig:fidelity}. Panels (a, b) correspond to GaAs QDs and (c, d) to InAs QDs. Results for exciton and biexciton preparation are shown on the left (a, c) and right (b, d), respectively. The fidelity of preparing the exciton state behaves as one could predict. As the detuning increases, the overlap between spectral functions decreases because the evolution is driven by higher frequencies (the detuning range corresponds to $\omega_{\mathrm{ac}}$ between 1.29 and 3.36 meV), for which the phonon response is weaker; thus, the fidelity increases. The temperature increases the number of phonons that entangle with the qubit, increasing the protocol error. In the case of the biexciton preparation (the detuning range corresponds to $\omega_{\mathrm{ac}}$ between 2.91 and 2.07 meV), one of the main transition frequencies in the evolution is the one between dressed exciton and biexciton states. Thus, increasing the absolute value of the detuning causes this transition energy to decrease, increasing the overlap between spectral functions. On the other hand, decreasing the absolute value of this detuning below $2.2$~meV causes difficulties due to the mixing of the exciton and biexciton states.

\begin{figure}[tb!]
    \centering
    \includegraphics[width=0.95\columnwidth]{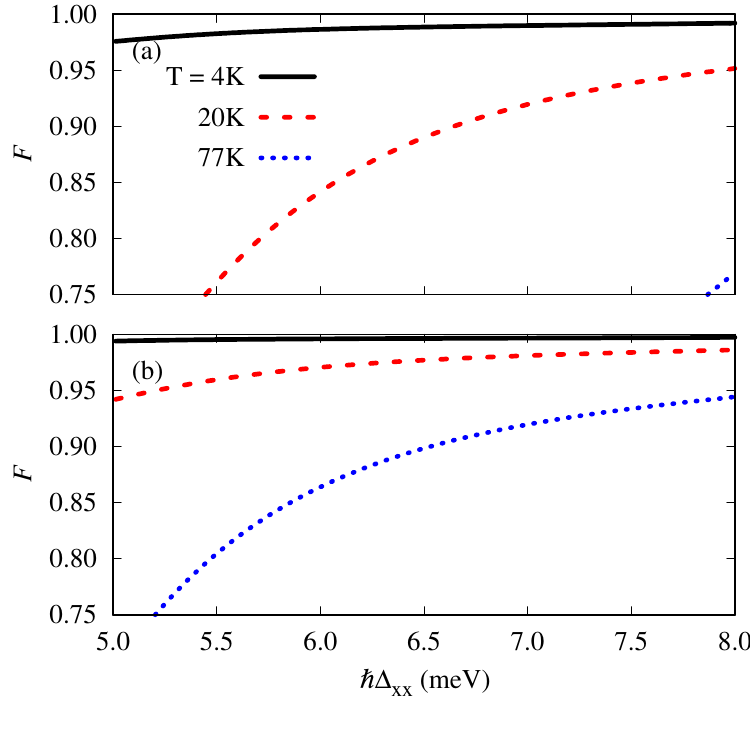}
    \caption{Achievable fidelity of the biexciton state preparation as a function of the laser detuning. The top row corresponds to InAs/GaAs QDs, and the bottom row corresponds to GaAs/AlGaAs QDs. The range of needed acoustic field energies is from $5.8$~meV to $8.2$~meV.}
    \label{fig:fidelityLargePhonons}
\end{figure}

\begin{figure}[tb!]
    \centering
    \includegraphics[width=.95\columnwidth]{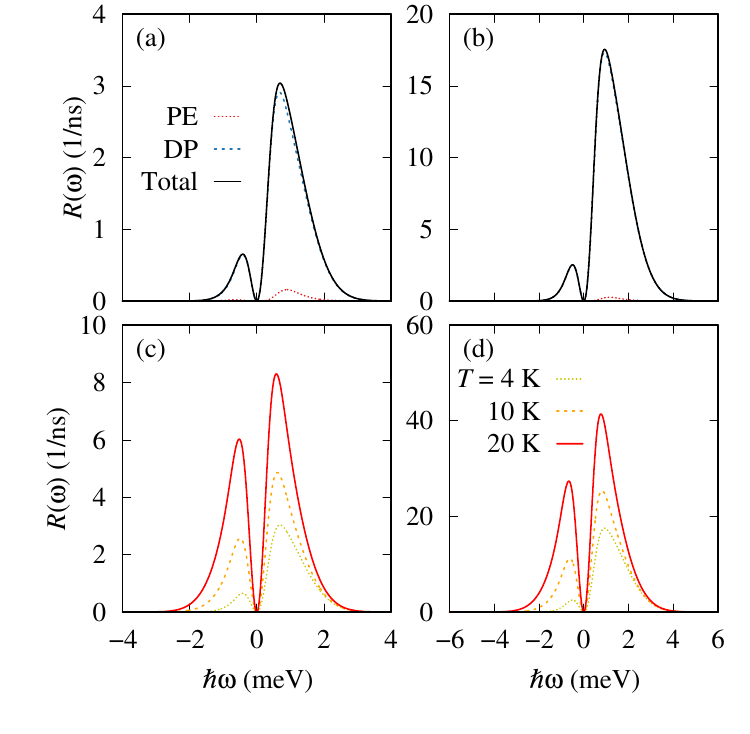}
    \caption{Phonon spectral densities: (a) Contribution of piezoelectric effect (PE), deformation potential (DP), and total spectral density for GaAs/AlGaAs QDs calculated at $T = 4$~K, (b) the same as (a) but for InAs/GaAs QDs, (c) Temperature dependence for GaAs/AlGaAs QDs, (d) Temperature dependence for InAs/GaAs QDs.}
    \label{fig:spectralDensity}
\end{figure}

As shown in Supplementary Figure~\ref{fig:fidelity}, the fidelity of the biexciton state preparation is significantly lower than for the exciton case for the sub-THz phonon frequencies studied here. We perform additional simulations to show that our biexciton state preparation can achieve a very high fidelity even at the liquid nitrogen temperature when terahertz acoustic frequencies are used. In Supplementary Figure~\ref{fig:fidelityLargePhonons}, we show the fidelity for larger detunings corresponding to driving with acoustic fields with the phonon energy in the $5.8$--$8.2$~meV range. In this case, we predict a significant increase in the fidelity for both types of QDs. For InAs QDs [\subfigref{fig:fidelityLargePhonons}{a}] it exceeds $95\%$ at $T=20$~K, and $75\%$ at $T=77$~K (fidelity can be further increased, increasing the detuning), while at lower detunings [Supplementary Figure~\ref{fig:fidelity}] the phonon impact was even beyond the perturbative regime. For GaAs QDs [\subfigref{fig:fidelityLargePhonons}{b}], we observe a qualitatively similar behavior of the fidelity as previously, but with even higher values. At $T=4$~K, the biexciton preparation is almost decoherence-free ($F>99.7\%$), and at higher temperatures, we also find significantly improved fidelity compared to the sub-THz acoustic modulation.

\subsection*{Phonon spectral densities}

Spectral densities for the two types of studied QDs and the contributions of PE and DP couplings at a low temperature of 4~K are shown in \subfigref{fig:spectralDensity}{a} and \subfigref{fig:spectralDensity}{b}. The temperature dependence is presented in \subfigref{fig:spectralDensity}{c} and \subfigref{fig:spectralDensity}{d}. \vspace{2em}

%